\documentclass[aps,prd,nofootinbib,showpacs,superscriptaddress, preprint]{revtex4-1}
\pdfoutput=1
\usepackage[utf8]{inputenc}
\usepackage{amsmath,amssymb}
\usepackage{epsfig}
\usepackage{graphicx}
\usepackage[usenames,dvipsnames]{color}
\usepackage{float}
\usepackage{bbold}
\usepackage[colorlinks,citecolor=blue]{hyperref}
\usepackage{color}
\usepackage{subfigure}
\begin{document}
%opening
\title{TeV Scale Leptogenesis with Dark Matter in \\ Non-standard Cosmology}

%%%%%%%%%   Authors   %%%%%%%%%%%%
\author{Devabrat Mahanta}
\email{devab176121007@iitg.ac.in}
\affiliation{Department of Physics, Indian Institute of Technology Guwahati, Assam 781039, India}

\author{Debasish Borah}
\email{dborah@iitg.ac.in}
\affiliation{Department of Physics, Indian Institute of Technology Guwahati, Assam 781039, India}

\begin{abstract}
We study the consequence of a non-standard cosmological epoch in the early universe on the generation of baryon asymmetry through leptogenesis as well as dark matter abundance. We consider two different non-standard epochs: one where a scalar field behaving like pressure-less matter dominates the early universe, known as early matter domination (EMD) scenario while in the second scenario, the energy density of the universe is dominated by a component whose energy density red-shifts faster than radiation, known as fast expanding universe (FEU) scenario. While a radiation dominated universe is reproduced by the big bang nucleosynthesis (BBN) epoch in both the scenario, the high scale phenomena like generation of baryon asymmetry and dark matter relic get significantly affected. Adopting a minimal particle physics framework known as the scotogenic model which generates light neutrino masses at one-loop level, we find that in one specific realisation of EMD scenario, the scale of leptogenesis can be lower than that in a standard cosmological scenario. The other non-standard cosmological scenarios, on the other hand, can be constrained from the requirement of successful low scale leptogenesis and generating correct dark matter abundance simultaneously. Such a low scale scenario not only gives a unified picture of baryon asymmetry, dark matter and origin of neutrino mass but also opens up interesting possibilities for experimental detection.
\end{abstract}
\maketitle
%\flushbottom

\section{Introduction}
Although the universe is supposed to start in a matter-antimatter symmetric manner, it is well known that the present universe has abundance of matter over antimatter. To explain this, one must find a dynamical mechanism to create an asymmetry such that after annihilation of matter with antimatter, a leftover matter component remains which gives rise to most of the visible matter we see around us. This observed excess of matter or baryons over antibaryons is often quantified in terms of baryon to photon ratio \cite{Tanabashi:2018oca, Aghanim:2018eyx}
\begin{equation}
\eta_B = \frac{n_{B}-n_{\bar{B}}}{n_{\gamma}} = 6.1 \times 10^{-10}.
\label{etaBobs}
\end{equation}
In order to generate this asymmetry dynamically, a well known prescription, known as Sakharov's conditions \cite{Sakharov:1967dj} must be realised in the early universe. These conditions are (i) baryon number (B) violation, (ii) C and CP violation and (iii) departure from thermal equilibrium. While all these conditions can not be realised in appropriate amount in the standard model (SM), several beyond standard model (BSM) proposals have been put forward to explain this observed baryon asymmetry of the universe (BAU) in a dynamical manner. The simplest of such mechanisms, perhaps, is to include additional heavy particles which can decay (or annihilate) into SM particles in a way which satisfies all the conditions mentioned above, leading to successful baryogenesis \cite{Weinberg:1979bt, Kolb:1979qa}. Another interesting way, which also connects the lepton sector physics, is known as leptogenesis which was proposed a few decades back \cite{Fukugita:1986hr}. For a review of leptogenesis, please see \cite{Davidson:2008bu}. In leptogenesis, instead of creating a baryon asymmetry directly from B violating interactions, an asymmetry in lepton sector is created via lepton number (L) violating processes (decay or scattering). If this lepton asymmetry is generated before the electroweak phase transition (EWPT), then the $(B+L)$-violating EW sphaleron transitions~\cite{Kuzmin:1985mm} can convert it to the required baryon asymmetry. Since the quark sector CP violation is insufficient to produce the required baryon asymmetry, the mechanism of leptogenesis can rely upon lepton sector CP violation which may be quite large as hinted by some neutrino oscillation experiments \cite{Esteban:2018azc}. An interesting feature of this scenario is that the required lepton asymmetry can be generated through CP violating out-of-equilibrium decays of the same heavy fields that take part in popular seesaw mechanisms~\cite{Minkowski:1977sc, Mohapatra:1979ia, Yanagida:1979as, GellMann:1980vs, Glashow:1979nm, Schechter:1980gr} that also explains the origin of tiny neutrino masses~\cite{Tanabashi:2018oca}, another observed phenomenon which the SM fails to address. While there are several different realisations of leptogenesis within particle physics frameworks, there have been serious efforts to realise it in a low scale BSM scenario so that it can be probed at different ongoing or near future experiments. For reviews of such low scale scenarios, please see.

Apart from the observed baryon asymmetry, there have been equally convincing evidences suggesting the presence of a mysterious, non-luminous form of matter, popularly known as dark matter (DM), in large amount in the present universe. In terms of density 
parameter $\Omega_{\rm DM}$ and $h = \text{Hubble Parameter}/(100 \;\text{km} ~\text{s}^{-1} 
\text{Mpc}^{-1})$, the present DM abundance is conventionally reported as \cite{Aghanim:2018eyx}:
%\begin{equation}
$\Omega_{\text{DM}} h^2 = 0.120\pm 0.001$
%\label{dm_relic}
%\end{equation}
at 68\% CL. In spite of astrophysics and cosmology based evidences \cite{Zwicky:1933gu, Rubin:1970zza, Clowe:2006eq, Aghanim:2018eyx, Tanabashi:2018oca}, the particle nature of DM is not yet known with none of the SM particles being a suitable DM candidate. Among the BSM proposals that have been put forward to solve the DM problem, the weakly interacting massive particle (WIMP) paradigm~\cite{Kolb:1990vq} is the most popular one. For a recent review of WIMP models, please see \cite{Arcadi:2017kky}. In this framework, a dark matter candidate typically having mass in the GeV-TeV scale and interaction rate similar to electroweak interactions can give rise to the correct dark matter relic abundance, a remarkable coincidence often referred to as the \textit{WIMP Miracle}. Such interactions enable the WIMP DM to be produced in thermal equilibrium in the early universe and eventually its number density gets frozen out when the rate of expansion of the universe takes over the interaction rates. Due to such sizeable interactions with the SM particles, WIMP type DM can leave observable signatures at direct detection experiments looking for DM-nucleon scatterings. However, no positive signal of this type have been observed by several direct detection experiments including LUX \cite{Akerib:2016vxi}, PandaX-II \cite{Tan:2016zwf, Cui:2017nnn} and Xenon1T \cite{Aprile:2017iyp, Aprile:2018dbl}.

One common feature of the BSM proposals for leptogenesis and WIMP DM is that both are high scale phenomena. While the generation of lepton asymmetry must occur before the electroweak scale $T \sim \mathcal{O}(100 \; \rm GeV)$, WIMP DM having mass $M_{\rm DM}$ usually freezes out at temperature $M_{\rm DM}/T \sim \mathcal{O}(20-30)$. Both these temperatures correspond to radiation dominated era of standard $\Lambda {\rm CDM}$ cosmology. However, there is no experimental evidence to suggest that the universe was radiation dominated prior to the era of the big bang nucleosynthesis (BBN) that is typically around 1 s after the big bang, corresponding to temperature of order $T \sim \mathcal{O}(1 \; \rm MeV)$. Since the predictions of leptogenesis and DM freeze-out crucially depends upon the rate of expansion of the universe, the standard predictions can change significantly if the universe expands at a rate different from usual radiation dominated one. This can happen in two different ways, one in which there was an early matter dominated (EMD) phase and the other in which the universe is dominated by a component that redshifts faster than radiation often dubbed as a fast expanding universe (FEU). There have been several works considering the DM relic abundance calculation in such non-standard cosmologies  \cite{McDonald:1989jd, Kamionkowski:1990ni, Chung:1998rq, Moroi:1999zb, Giudice:2000ex, Allahverdi:2002nb, Allahverdi:2002pu, Acharya:2009zt, Davoudiasl:2015vba, Drees:2018dsj, Bernal:2018ins, Bernal:2018kcw, Arias:2019uol, Delos:2019dyh, Chanda:2019xyl, Bernal:2019mhf, Poulin:2019omz, Maldonado:2019qmp, Betancur:2018xtj, DEramo:2017gpl, DEramo:2017ecx, Biswas:2018iny, Visinelli:2015eka, Visinelli:2017qga}. In this work, we study the impact of such non-standard cosmological histories on the genesis of baryon asymmetry via leptogenesis along with DM. As a concrete example, we consider a very minimal BSM framework where the particle content of the SM is extended by three copies of singlet right handed neutrinos and a scalar field doublet under the $SU(2)_L$ group of SM. All these additional fields are odd under an in-built and unbroken $Z_2$ symmetry while the SM fields are $Z_2$ even. It is the minimal model belonging to the scotogenic framework proposed by Ma in 2006 \cite{Ma:2006km}. While the lightest $Z_2$ odd particle is naturally stable and, if electromagnetically neutral, provides a viable DM candidate, light neutrino masses arise radiatively with the $Z_2$ particles going inside the loop. Apart from this, the out-of-equilibrium decay of the heavy singlet fermions can generate the required lepton asymmetry, which can give rise to the observed BAU after electroweak sphaleron transitions. Recently the authors of \cite{Hugle:2018qbw, Borah:2018rca} studied the possibility of creating lepton asymmetry from the decay of lightest singlet fermion $(N_1)$ decay and found that the required asymmetry can be produced for $M_1 \sim 10$ TeV within a vanilla leptogenesis framework having hierarchical $Z_2$ odd singlet fermionic masses while satisfying the constraints from light neutrino masses\footnote{Note that this is a significant improvement over the usual Davidson-Ibarra bound $M_1 > 10^9$ GeV for vanilla leptogenesis in type I seesaw framework \cite{Davidson:2002qv}}. Earlier works reporting low scale leptogenesis with $M_1$ having mass around few tens of TeV can be found in \cite{Hambye:2009pw, Racker:2013lua} where the author considered a hierarchical spectrum of right handed neutrinos. TeV scale leptogenesis in this model with quasi-degenerate right handed neutrinos was also discussed in earlier works \cite{Kashiwase:2012xd, Kashiwase:2013uy}. A high scale leptogenesis version of this scenario was studied by the authors of \cite{Huang:2018vcr}. An alternate possibility with keV scale $N_1$ DM and leptogenesis from a combination of the Akhmedov-Rubakov-Smirnov (ARS) mechanism \cite{Akhmedov:1998qx} and scalar doublet decay \cite{Hambye:2016sby} was recently explored by the authors of \cite{Baumholzer:2018sfb}. Recently, the possibility of singlet fermion DM in this model was studied where BAU is generated from the decay of heavier singlet neutrinos \cite{Mahanta:2019gfe}. To summarise, it is possible to realise successful leptogenesis at around 10 TeV while satisfying the bounds from neutrino mass and DM relic abundance in this model. In this work, our goal is to check if this scale of leptogenesis can be further lowered with a non-standard cosmological epoch in the early universe. While finalising this work, we noticed this recent work \cite{Chen:2019etb} where authors studied high scale type I seesaw leptogenesis in a fast expanding universe. A relatively earlier work \cite{Abdallah:2012nm} considered the effects of non-standard cosmology (braneworld scenario and non-thermal production of right handed neutrinos from inflaton decay) on generation of lepton asymmetry in a TeV scale inverse seesaw model. Low scale leptogenesis in scalar-tensor theories of gravity was studied in \cite{Dutta:2018zkg} while the implications of such non-standard cosmology on DM relic were studied in several earlier works including \cite{Catena:2004ba, Catena:2009tm, Meehan:2015cna, Dutta:2016htz, Dutta:2017fcn}. Our work is different from both these two works due to the non-standard cosmological scenarios considered, different way of generating neutrino mass as well as connection to dark matter. We find that for one specific realisation of EMD era, it is indeed possible to lower the scale of leptogenesis to a few TeV while in agreement with neutrino and DM related constraints. Such a low scale scenario can have tantalising prospects of detection at different experiments, which we briefly comment upon. At the same time, we also constrain such non-standard cosmological scenarios from the requirement of successful leptogenesis.

This paper is organised as follows. In section \ref{sec2}, we discuss the model and summarise the details of dark matter and leptogenesis calculations in a standard cosmological scenario. In section \ref{sec:EMD} we discuss the generation of baryon asymmetry and DM relic in an early matter dominated scenario. The same discussion is extended to an another non-standard cosmological scenario called fast expanding universe in section \ref{sec:feu}. We briefly discuss the leptogenesis results with the inclusion of lepton flavour effects in section \ref{sec:flav} and finally conclude in section \ref{sec:conc}.

\section{Scotogenic Model}
\label{sec2}
As pointed out earlier, we adopt a specific particle physics model in order to show the effects of non-standard cosmology on the origin of baryon asymmetry. This is also the the minimal model belonging to the scotogenic framework where the SM is extended by three copies of SM-singlet fermions $N_i$ (with $i=1,2,3$) and one $SU(2)_L$-doublet scalar field $\eta$ (also called inert doublet), all being odd under an in-built and unbroken $Z_2$ symmetry. On the other hand, the SM fields remain $Z_2$-even, i.e. under the $Z_2$-symmetry, we have  
\begin{align}
N_i \rightarrow -N_i, \quad  \eta  \rightarrow -\eta, \quad \Phi_{1} \rightarrow \Phi_{1}, \quad \Psi_{\rm SM} \to 
\Psi_{\rm SM} \, ,
\label{eq:Z2}
\end{align}
where $\Phi_1$ is the SM Higgs doublet and $\Psi_{\rm SM}$'s stand for the SM fermions. This $Z_2$ symmetry, though \textit{ad hoc} in this minimal setup, could be realised naturally as a subgroup of a continuous gauge symmetry like $U(1)_{B-L}$ with non-minimal field content \cite{Dasgupta:2014hha,Das:2017ski}. 
%The unbroken $Z_2$ symmetry also forbids the second Higgs doublet or the inert doublet to acquire any non-zero vacuum expectation value (VEV). The $Z_2$ odd nature ensures that the second Higgs doublet couples to lepton doublets only via interactions involving the $Z_2$ odd singlet fermions. Since the interactions of the second Higgs doublet with the usual SM fermions are forbidden at renormalisable level, it is often referred to as the inert doublet. 

The leptonic Yukawa Lagrangian is
\begin{equation}\label{IRHYukawa}
{\cal L} \ \supset \ \frac{1}{2}(M_N)_{ij} N_iN_j + \left(Y_{ij} \, \bar{L}_i \tilde{\eta} N_j  + \text{h.c.} \right) \ . 
\end{equation}
The $Z_2$ symmetry forbids the generation of light neutrino masses at tree level through the conventional type I seesaw mechanism \cite{Minkowski:1977sc, Mohapatra:1979ia, Yanagida:1979as, GellMann:1980vs, Glashow:1979nm, Schechter:1980gr} by preventing the usual Dirac Yukawa term $\bar{L}\tilde{\Phi}_1 N$ involving the SM Higgs. The scalar sector of the model resembles the one in the inert Higgs doublet model (IHDM) \cite{Deshpande:1977rw}, a minimal extension of the SM by a $Z_2$ odd scalar doublet $\eta$~\cite{Ma:2006km, Dasgupta:2014hha, Cirelli:2005uq, Barbieri:2006dq, Ma:2006fn,  LopezHonorez:2006gr,  Hambye:2009pw, Dolle:2009fn, Honorez:2010re, LopezHonorez:2010tb, Gustafsson:2012aj, Goudelis:2013uca, Arhrib:2013ela, Diaz:2015pyv, Ahriche:2017iar}. 
%The $Z_2$ symmetry prevents linear and trilinear terms of the inert doublet with the SM Higgs. The bare mass squared term of the inert doublet is chosen to be positive definite in order to ensure that it does not acquire any non-zero VEV. Absence of linear terms ensures that it does not even acquire any induced VEV after electroweak symmetry breaking (EWSB). 
The scalar potential of the model can be written as
\begin{align}
V(\Phi_1,\eta) & \ = \   \mu_1^2|\Phi_1|^2 +\mu_2^2|\eta|^2+\frac{\lambda_1}{2}|\Phi_1|^4+\frac{\lambda_2}{2}|\eta|^4+\lambda_3|\Phi_1|^2|\eta|^2 \nonumber \\
& \qquad +\lambda_4|\Phi_1^\dag \eta|^2 + \left[\frac{\lambda_5}{2}(\Phi_1^\dag \eta)^2 + \text{h.c.}\right] \, . \label {c}
\end{align}
where $\Phi_1$ is the SM Higgs doublet. In order to ensure that none of the neutral components of the inert Higgs doublet $\eta$ acquire a nonzero VEV, $\mu_2^2 >0$ is assumed. This also ensures that the $Z_2$ symmetry does not get spontaneously broken, leaving the lightest $Z_2$ odd particle stable and hence, if electromagnetically neutral, a suitable DM candidate. The electroweak symmetry breaking (EWSB) occurs due to the nonzero VEV acquired by the neutral component of SM like Higgs doublet $\Phi_1$. 

After the EWSB, these two scalar doublets can be written in the following form (assuming unitary gauge):
\begin{equation}
\Phi_1 \ = \ \begin{pmatrix} 0 \\  \frac{ v +h }{\sqrt 2} \end{pmatrix} , \qquad \eta \ = \ \begin{pmatrix} H^\pm\\  \frac{H^0+iA^0}{\sqrt 2} \end{pmatrix} \, ,
\label{eq:idm}
\end{equation}
where $h$ is the SM-like Higgs boson, $H^0$ and $A^0$ are the CP-even and CP-odd scalars, and $H^\pm$ are the charged scalars from the inert Higgs doublet. Using these notations, the physical scalar masses can be written as \begin{eqnarray}
m_h^2 & \ = \ & \lambda_1 v^2 ,\nonumber\\
m_{H^\pm}^2 & \ = \ & \mu_2^2 + \frac{1}{2}\lambda_3 v^2 , \nonumber\\
m_{H^0}^2 & \ = \ & \mu_2^2 + \frac{1}{2}(\lambda_3+\lambda_4+\lambda_5)v^2 \ = \ m^2_{H^\pm}+
\frac{1}{2}\left(\lambda_4+\lambda_5\right)v^2  , \nonumber\\
m_{A^0}^2 & \ = \ & \mu_2^2 + \frac{1}{2}(\lambda_3+\lambda_4-\lambda_5)v^2 \ = \ m^2_{H^\pm}+
\frac{1}{2}\left(\lambda_4-\lambda_5\right)v^2 \, .
\label{mass_relation}
\end{eqnarray}
In our study, we consider $ \lambda_5 >0$ which implies that the CP-odd scalar is lighter than the CP-even one. This corresponds to $\lambda_L= (\lambda_3+\lambda_4-\lambda_5)$ being the DM-Higgs coupling.
%\begin{figure}[t!]
%\centering
%\includegraphics[scale=0.5]{Neumass-2-eps-converted-to.pdf}
%\caption{One-loop contribution to neutrino mass in the scotogenic model.}
%\label{fig1a}
%\end{figure}

Light neutrino masses which arise at one loop level can be evaluated as ~\cite{Ma:2006km, Merle:2015ica}
\begin{align}
(M_{\nu})_{ij} \ & = \ \sum_k \frac{Y_{ik}Y_{jk} M_{k}}{32 \pi^2} \left ( \frac{m^2_{H^0}}{m^2_{H^0}-M^2_k} \: \text{ln} \frac{m^2_{H^0}}{M^2_k}-\frac{m^2_{A^0}}{m^2_{A^0}-M^2_k}\: \text{ln} \frac{m^2_{A^0}}{M^2_k} \right) \nonumber \\ 
& \ \equiv  \ \sum_k \frac{Y_{ik}Y_{jk} M_{k}}{32 \pi^2} \left[L_k(m^2_{H^0})-L_k(m^2_{A^0})\right] \, ,
\label{numass1}
\end{align}
where %$m^2_{R,I}=m^2_{H,A}$ are the masses of scalar and pseudo-scalar part of $\Phi^0_2$ and 
$M_k$ is the mass eigenvalue of the mass eigenstate $N_k$ in the internal line and the indices $i, j = 1,2,3$ run over the three neutrino generations as well as three copies of $N_i$. The function $L_k(m^2)$ is defined as 
\begin{align}
L_k(m^2) \ = \ \frac{m^2}{m^2-M^2_k} \: \text{ln} \frac{m^2}{M^2_k} \, .
\label{eq:Lk}
\end{align}
From the physical scalar mass expressions given in equations \eqref{mass_relation}, one can write $m^2_{H^0}-m^2_{A^0}=\lambda_5 v^2$. Thus, light neutrino mass is directly proportional to the parameter $\lambda_5$. In fact, the $\lambda_5$-term in the scalar potential~\eqref{c} breaks lepton number by two units, when considered together with the SM-singlet fermions Lagrangian~\eqref{IRHYukawa}. Therefore, in addition to the one loop suppression factor and the Dirac Yukawa couplings, one has more freedom in tuning this quartic coupling $\lambda_5$ in order to generate the required sub-eV light neutrino mass even with TeV scale right handed neutrinos. Since setting $\lambda_5 \to 0$ allows us to recover the lepton number global symmetry, the smallness of $\lambda_5$ is technically natural in the 't Hooft sense~\cite{tHooft:1979rat}. Also, $\lambda_5$ decides the mass splitting between $A^0, H^0$ which can be constrained by dark matter direct detection limits, as studied earlier by several authors \cite{Dasgupta:2014hha, LopezHonorez:2006gr,  Hambye:2009pw, Dolle:2009fn, Honorez:2010re, LopezHonorez:2010tb, Gustafsson:2012aj, Goudelis:2013uca, Arhrib:2013ela, Diaz:2015pyv, Ahriche:2017iar}.

Since the model we adopt here also provides a solution to the neutrino mass problem as discussed above, it is important to ensure that the choices of Dirac Yukawa couplings as well as other parameters are consistent with the cosmological upper bound on the sum of neutrino masses, $\sum_i m_{i}\leq 0.11$ eV~\cite{Aghanim:2018eyx}, as well as the neutrino oscillation data on three mixing angles and two mass squared differences~\cite{deSalas:2017kay, Esteban:2018azc}. In order to incorporate these bounds from light neutrino sector, it is convenient to write the one loop neutrino mass formula \eqref{numass1} in a form similar to that of type I seesaw: 
\begin{align}
M_\nu \ = \ Y {\Lambda}^{-1} Y^T \, ,
\label{eq:nu2}
\end{align}
where, $\Lambda$ is a diagonal matrix with elements given by 
\begin{align}
 \Lambda_i \ & = \ \frac{2\pi^2}{\lambda_5}\zeta_i\frac{2M_i}{v^2} \, , \\
\textrm {and}\quad \zeta_i & \ = \  \left(\frac{M_{i}^2}{8(m_{H^0}^2-m_{A^0}^2)}\left[L_i(m_{H^0}^2)-L_i(m_{A^0}^2) \right]\right)^{-1} \, . \label{eq:zeta}
\end{align}
The light neutrino mass matrix~\eqref{eq:nu2} which is complex symmetric by virtue of its Majorana nature, can be diagonalised by the usual Pontecorvo-Maki-Nakagawa-Sakata (PMNS) mixing matrix $U$ (in the diagonal charged lepton basis), written in terms of neutrino oscillation data (up to the Majorana phases) as
\begin{equation}
U=\left(\begin{array}{ccc}
c_{12}c_{13}& s_{12}c_{13}& s_{13}e^{-i\delta}\\
-s_{12}c_{23}-c_{12}s_{23}s_{13}e^{i\delta}& c_{12}c_{23}-s_{12}s_{23}s_{13}e^{i\delta} & s_{23}c_{13} \\
s_{12}s_{23}-c_{12}c_{23}s_{13}e^{i\delta} & -c_{12}s_{23}-s_{12}c_{23}s_{13}e^{i\delta}& c_{23}c_{13}
\end{array}\right) U_{\text{Maj}}
\label{PMNS}
\end{equation}
where $c_{ij} = \cos{\theta_{ij}}, \; s_{ij} = \sin{\theta_{ij}}$ and $\delta$ is the leptonic Dirac CP phase. The diagonal matrix $U_{\text{Maj}}=\text{diag}(1, e^{i\alpha}, e^{i(\zeta+\delta)})$ contains the undetermined Majorana CP phases $\alpha, \zeta$. The diagonal light neutrino mass matrix is therefore,
\begin{align}
D_\nu \ = \ U^\dag M_\nu U^* \ = \ \textrm{diag}(m_1,m_2,m_3) \, .
\end{align}   
where the light neutrino masses can follow either normal ordering (NO) or inverted ordering (IO). Without losing generality, we consider NO of light neutrino masses for our numerical analysis. Since the inputs from neutrino data are only in terms of the mass squared differences and mixing angles, it would be  
useful for our purpose to express the Yukawa couplings in terms of light neutrino parameters. This is possible through the Casas-Ibarra (CI) parametrisation \cite{Casas:2001sr} extended to radiative seesaw model \cite{Toma:2013zsa} which allows us to write the Yukawa coupling matrix satisfying the neutrino data as
\begin{align}
Y \ = \ U D_\nu^{1/2} R^{\dagger} \Lambda^{1/2} \, ,
\label{eq:Yuk}
\end{align}
where $R$ is an arbitrary complex orthogonal matrix satisfying $RR^{T}=\mathbb{1}$.

\subsection{Dark matter}
%\label{sec3}
As pointed out earlier, the DM candidate in our model is one of the neutral components of the $Z_2$ odd scalar doublet $\eta$. By virtue of its SM gauge interactions, DM can be thermally produced in the early universe and hence give rise to a WIMP type scenario. Apart from gauge interactions, the Higgs portal interactions can also play a non-trivial role in generating thermal relic abundance.

For WIMP type DM which is produced thermally in the early universe, its thermal relic abundance can be obtained by solving the Boltzmann equation for the evolution of the DM number density $n_{\rm DM}$:
\begin{equation}
\frac{dn_{\rm DM}}{dt}+3Hn_{\rm DM} \ = \ -\langle \sigma v \rangle \left[n^2_{\rm DM} -(n^{\rm eq}_{\rm DM})^2\right],
\label{eq:BE}
\end{equation}
where $n^{\rm eq}_{\rm DM}$ is the equilibrium number density of DM and $ \langle \sigma v \rangle $ is the thermally averaged annihilation cross section, given by~\cite{Gondolo:1990dk}
\begin{equation}
\langle \sigma v \rangle \ = \ \frac{1}{8m_{\rm DM}^4T K^2_2\left(\frac{m_{\rm DM}}{T}\right)} \int\limits^{\infty}_{4m_{\rm DM}^2}\sigma (s-4m_{\rm DM}^2)\sqrt{s}\: K_1\left(\frac{\sqrt{s}}{T}\right) ds \, ,
\label{eq:sigmav}
\end{equation}
where $K_i(z)$'s are modified Bessel functions of order $i$. In the presence of coannihilation, one follows the recipe given by~\cite{Griest:1990kh} to calculate the relic abundance. Apart from the gauge mediated coannihilations, there can be coannihilations due to lepton portal interactions as well, if the mass of $\eta$ remain very close to that of $N_1$. We have used \texttt{micrOMEGAs} package \cite{Belanger:2013oya} to calculate the freeze-out details of DM in our work.

\subsection{Leptogenesis}
%\label{sec4}
Successful leptogenesis is possible in this model due to the presence of heavy singlet neutrinos $N_i$ whose out-of-equilibrium decay into SM leptons and $\eta$ can generate the required non-zero lepton asymmetry~\cite{Ma:2006fn, Kashiwase:2012xd, Kashiwase:2013uy, Racker:2013lua, Clarke:2015hta, Hugle:2018qbw, Borah:2018rca}. In the hierarchical spectrum of $N_i, i=1-3$ one can significantly lower the usual Davidson-Ibarra bound to around 10 TeV \cite{Hugle:2018qbw, Borah:2018rca} without any need of resonance enhancement \cite{Pilaftsis:2003gt, Dev:2017wwc}. Although $N_{2,3}$ decay can also generate lepton asymmetry, in principle, we consider the asymmetry generated by $N_{2,3}$ decay or any pre-existing asymmetry to be negligible due to strong washout effects mediated either by $N_{1}$ or $N_{2,3}$ themselves.

The CP asymmetry parameter is defined as
\begin{equation}
\epsilon_{i} =\frac{\sum_{\alpha}\Gamma(N_{i}\rightarrow l_{\alpha}\eta)-\Gamma(N_{i}\rightarrow\bar{l_{\alpha}}\bar{\eta})}{\sum_{\alpha}\Gamma(N_{i}\rightarrow l_{\alpha}\eta)+\Gamma(N_{i}\rightarrow\bar{l_{\alpha}}\bar{\eta})}.
\label{epsilon1}
\end{equation} 
\\
The CP asymmetry parameter for $N_i \rightarrow l_{\alpha} \eta, \bar{l_{\alpha}}\bar{\eta}$ is given by 
\begin{equation}
\epsilon_{i \alpha} = \frac{1}{8 \pi (Y^{\dagger}Y)_{ii}} \sum_{j\neq i} \bigg [ f \left( \frac{M^2_j}{M^2_i}, \frac{m^2_{\eta}}{M^2_i} \right) {\rm Im} [ Y^*_{\alpha i} Y_{\alpha j} (Y^{\dagger} Y)_{ij}] - \frac{M^2_i}{M^2_j-M^2_i} \left( 1-\frac{m^2_{\eta}}{M^2_i} \right)^2 {\rm Im}[Y^*_{\alpha i} Y_{\alpha j} H_{ij}] \bigg ]
\label{epsilonflav}
\end{equation}
where, the function $f(r_{ji},\eta_{i})$ is coming from the interference of the tree-level and one loop diagrams and has the form
\begin{equation}
f(r_{ji},\eta_{i})=\sqrt{r_{ji}}\left[1+\frac{(1-2\eta_{i}+r_{ji})}{(1-\eta_{i}^{2})^{2}}{\rm ln}(\frac{r_{ji}-\eta_{i}^{2}}{1-2\eta_{i}+r_{ji}})\right]
\end{equation}
with $r_{ji}=M_{j}^{2}/M_{i}^{2}$ and $\eta_{i}=m_{\eta}^{2}/M_{i}^{2}$. The self energy contribution $H_{ij}$ is given by 
\begin{equation}
H_{ij} = (Y^{\dagger} Y)_{ij} \frac{M_j}{M_i} + (Y^{\dagger} Y)^*_{ij}
\end{equation}
Now, the CP asymmetry parameter, neglecting the flavour effects (summing over final state flavours $\alpha$) is
\begin{equation}
\epsilon_{i}=\frac{1}{8\pi(Y^{\dagger}Y)_{ii}}\sum_{j\neq i}{\rm Im}[((Y^{\dagger}Y)_{ij})^{2}]\frac{1}{\sqrt{r_{ji}}}F(r_{ji},\eta_{i})
 \label{eq:14}
\end{equation}
\\
where the function $F(r_{ji},\eta)$ is defined as 
\begin{equation}
F(r_{ji},\eta_{i})=\sqrt{r_{ji}}\left[ f(r_{ji},\eta_{i})-\frac{\sqrt{r_{ji}}}{r_{ji}-1}(1-\eta_{i})^{2} \right].
\end{equation}
Let us define the decay parameter as
\begin{equation}
K_{N_1}=\dfrac{\Gamma_{1}}{H(z=1)}
\end{equation}
where $\Gamma_{1}$ is the $N_{1}$ decay width, $H$ is the Hubble parameter and $z=M_{1}/T$ with $T$ being the temperature of the thermal bath. Leptogenesis occurs far above the electroweak scale where the universe was radiation dominated. In this era the Hubble parameter can be expressed in terms of the temperature $T$ as follows
\begin{equation}
H=\sqrt{\dfrac{8\pi^{3} g_{*}}{90}}\dfrac{T^{2}}{M_{Pl}}=H(z=1)\dfrac{1}{z^{2}}
\end{equation}
where $g_{*}$ is the effective number of relativistic degrees of freedom and $M_{\rm Pl}\simeq 1.22\times10^{19}$ GeV is the Planck mass. The decay width $\Gamma_{1}$ can be calculated as 
\begin{equation}
\Gamma_{1}=\dfrac{M_{2}}{8\pi}(Y^{\dagger} Y)_{11}(1-\eta_{1})^{2}
\end{equation}
The frequently appearing $Y^{\dagger}Y$ is calculated using Casas-Ibarra parametrisation mentioned earlier as 
\begin{equation}
(Y^{\dagger}Y)_{ij}=\sqrt{\Lambda_{i}\Lambda_{j}}(RD_{\nu}R^{\dagger})_{ij}
\label{eq:20}
\end{equation}
$D_{\nu}={\rm diag}(m_{1},m_{2},m_{3})$ is the diagonal light neutrino mass matrix. One important point here is to note down that the important quantity $Y^{\dagger}Y$ for leptogenesis is independent of the lepton mixing PMNS matrix, whereas it is dependent on the complex angles of the CI parametrisation. Thus the CP violating phases relevant for leptogenesis are independent of the CP violating phases in the PMNS matrix. The dependence of the CP asymmetry on $M_{i}$ and $\lambda_{5}$ is evident through $\Lambda_{i}$.

The Boltzmann equations for leptogenesis are given by~\cite{Buchmuller:2004nz}
\begin{eqnarray}
\frac{dn_{N_1}}{dz}& \ = \ &-D_1 (n_{N_1}-n_{N_1}^{\rm eq}) \, , \label{eq:bol1} \\
\frac{dn_{B-L}}{dz}& \ = \ &-\epsilon_1 D_1 (n_{N_1}-n_{N_1}^{\rm eq})-W^{\rm Total}n_{B-L} \, , \label{eq:bol2}
\end{eqnarray}
where $n_{N_1}^{\rm eq}=\frac{z^2}{2}K_2(z)$ is the equilibrium number density of $N_1$ (with $K_i(z)$ being the modified Bessel function of $i$-th kind). The quantity on the right hand side of the above equations
\begin{align}
D_1 \ \equiv \ \frac{\Gamma_{1}}{Hz} \ = \ K_{N_1} z \frac{K_1(z)}{K_2(z)}
\end{align}
 measures the total decay rate of $N_1$ with respect to the Hubble expansion rate, and similarly, $W^{\rm Total} \equiv \frac{\Gamma_{W}}{Hz}$ measures the total washout rate. The washout term is the sum of two contributions, i.e. $W^{\rm Total}=W_1+W_{\Delta L}$, where the washout due to the inverse decays $\ell \eta$, $\bar\ell \eta^* \rightarrow N_1$ is given by 
\begin{align}
W_1=W_{\rm ID} \ = \ \frac{1}{4}K_{N_1} z^3 K_1(z).
\end{align}
The other contribution to washout $ W_{\Delta L}$ originates from scatterings which violate lepton number by $\Delta L=1, 2$. The contribution from $\Delta L=2$ scatterings $\ell \eta \leftrightarrow \bar{\ell} \eta^*$, $\ell\ell\leftrightarrow \eta^* \eta^*$ is given by~\cite{Hugle:2018qbw} 
\begin{align}
W_{\Delta L=2} \ \simeq \ \frac{18\sqrt{10}\,M_{\rm Pl}}{\pi^4 g_\ell \sqrt{g_*}z^2 v^4}\left(\frac{2\pi^2}{\lambda_5}\right)^2 M_{1}\bar{m}_\zeta^2 \, ,
\label{eq:wash2}
\end{align}
where we have assumed $\eta_1\ll 1$ for simplicity, $g_\ell$ stands for the internal degrees of freedom for the SM leptons, and  $\bar{m}_\zeta$ is the effective neutrino mass parameter, defined as 
\begin{align}
\bar{m}_\zeta^2 \  \simeq \  4\zeta_1^2 m_{l}^2+\zeta_2 m_{h_1}^2+\zeta_3^2 m_{h_2}^2 \, ,
\end{align}
with $m_l, m_{h_1, h_2}$ are being the lightest and heavier neutrino mass eigenvalues, $\zeta_i$ defined in equation~\eqref{eq:zeta} and $L_i(m^2)$ defined in equation~\eqref{eq:Lk}. It should be noted that equation~\eqref{eq:wash2} is similar to the $\Delta L=2$ washout term in vanilla leptogenesis, except for the $\left(\frac{2\pi^2}{\lambda_5}\right)^2$ factor.

Upon solving the above Boltzmann equations \eqref{eq:bol1} and \eqref{eq:bol2} simultaneously, the final $B-L$ asymmetry $n_{B-L}^f$ just before electroweak sphaleron freeze-out is converted into the observed baryon to photon ratio by the standard formula 
\begin{align}
\eta_B \ = \ \frac{3}{4}\frac{g_*^{0}}{g_*}a_{\rm sph}n_{B-L}^f \ \simeq \ 9.2\times 10^{-3}\: n_{B-L}^f \, ,
\label{eq:etaB}
\end{align}
where $a_{\rm sph}=\frac{8}{23}$ is the sphaleron conversion factor (taking into account two Higgs doublets). The effective relativistic degrees of freedom is taken to be $g_*=110.75$, slightly higher than that of the SM at such temperatures as we are including the contribution of the inert Higgs doublet too. In the above expression $g_*^0=\frac{43}{11}$ is the effective relativistic degrees of freedom at the recombination epoch.

\begin{figure}[h]
\begin{center}
\includegraphics[scale=.22]{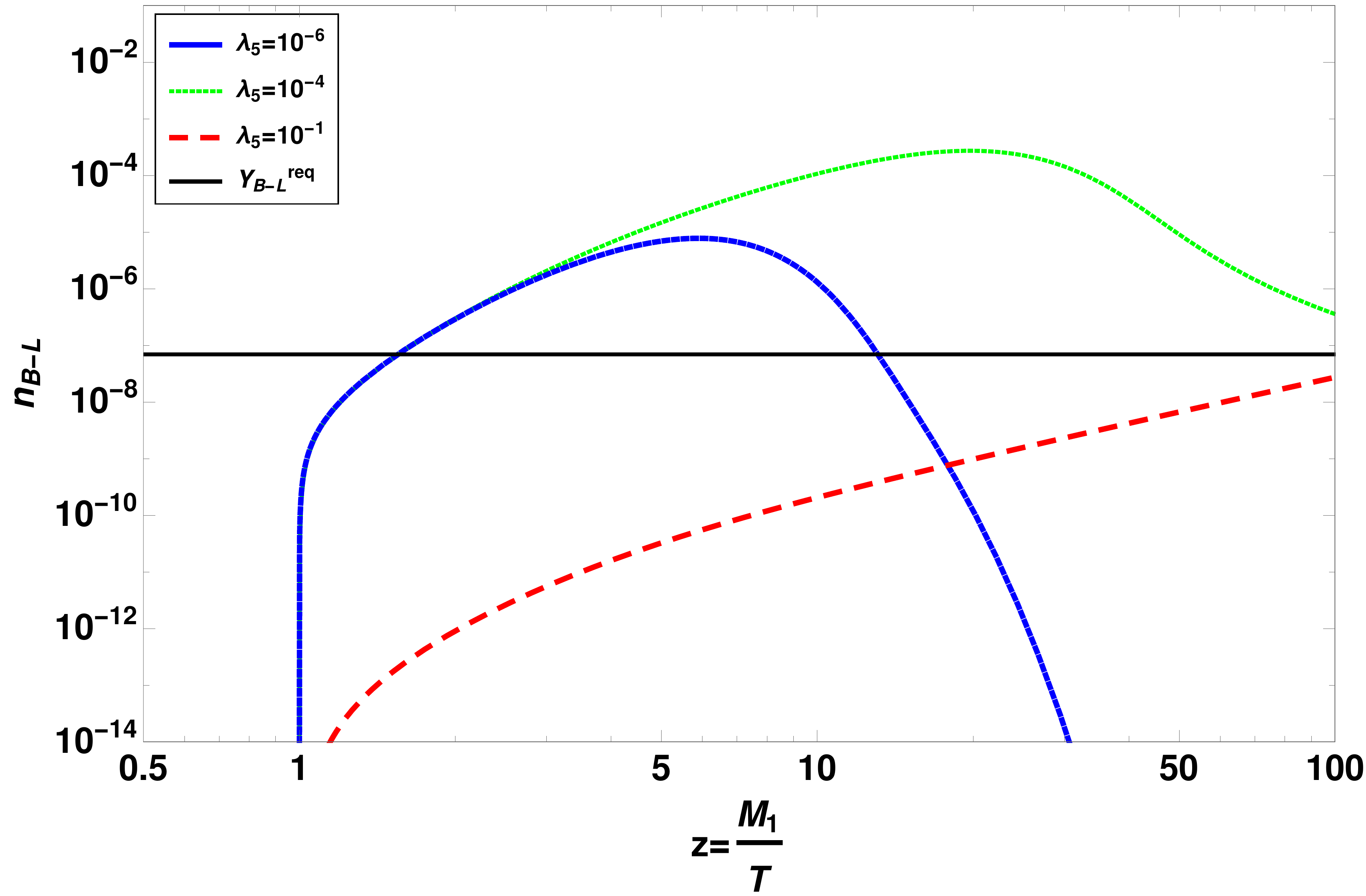}
\includegraphics[scale=.22]{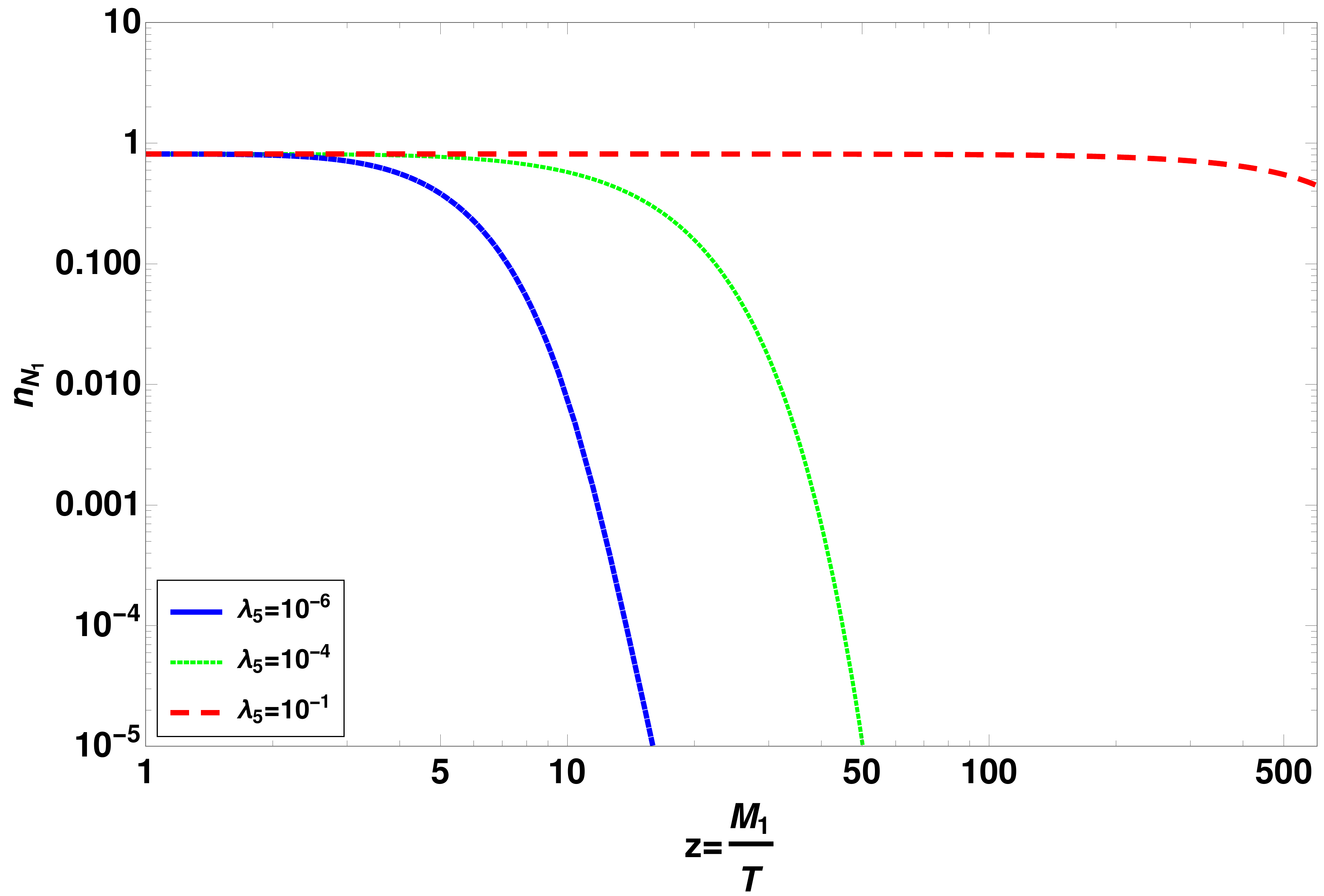}
\caption{Comoving number densities of $B-L$ (left panel) and ${N_{1}}$ (right panel) respectively for different benchmark parameters. The parameters used for this results are $M_{1}=10$ TeV, $m_{1}=10^{-13}$ eV, $\mu_2=100$ GeV and $M_{i+1}/M_{i}=10^{0.5}$.}
\label{standardlepto1}
\end{center}
\end{figure}

\begin{figure}[h]
\begin{center}
\includegraphics[scale=.25]{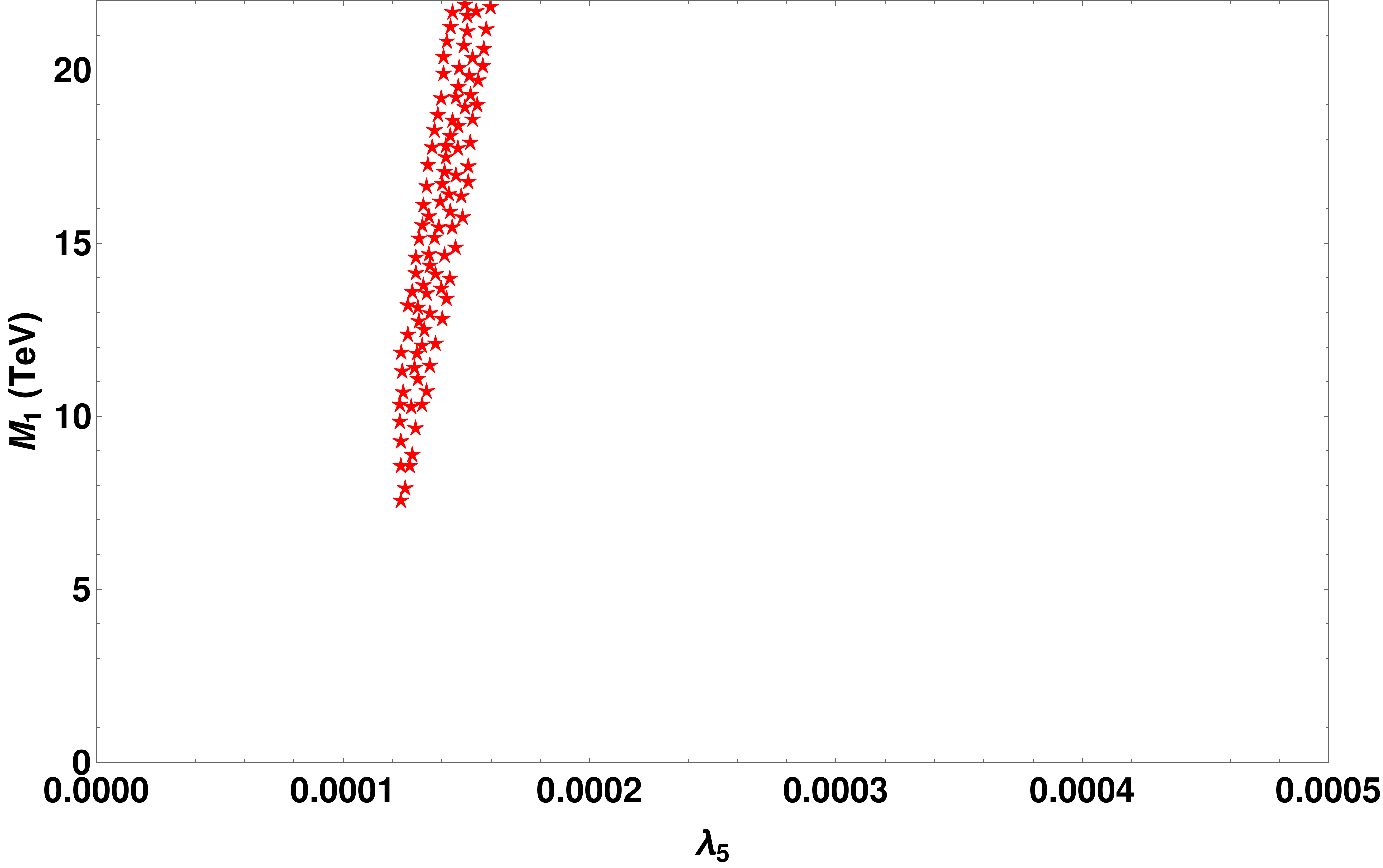}
\caption{Parameter space in the $M_{1}-\lambda_{5}$ plane that gives rise to observed baryon asymmetry in the standard radiation dominated universe. The parameters used for this results are $m_{1}=10^{-13}$ eV, $\mu_2=100$ GeV, $M_{i+1}/M_{i}=10^{0.5}$.}
\label{standardlepto2}
\end{center}
\end{figure}

Before studying the changes in leptogenesis results due to non-standard cosmological history, we first solve the above Boltzmann equations assuming a standard radiation dominated epoch. While more details can be found in earlier works \cite{Hugle:2018qbw, Borah:2018rca}, we show the evolution of lepton asymmetry and number density of $N_1$ in figure \ref{standardlepto1} for three different values of $\lambda_5$. As the number density of $N_1$ decreases due to its decay, the lepton asymmetry grows. The decrease in lepton asymmetry subsequently due to washout effects are clearly visible from left panel plot of figure \ref{standardlepto1}, where the parameter $\lambda_5$ plays a crucial role. 
For this as well as remaining calculations, the R matrix is chosen to have the following structure 
\begin{equation}
R=\begin{pmatrix}
\cos{z} & 0 & \sin{z}&\\
0  & 1 & 0 \\
 -\sin{z} & 0 & \cos{z} 
\end{pmatrix}
\end{equation}
where $z$ is a complex angle with, $z_{R}=z_{I}=\sqrt{\dfrac{m_{1}}{2m_{3}}}$ \cite{Hugle:2018qbw}. We also consider normal ordering of light neutrino mass with vanishingly small lightest neutrino mass $m_1 = 10^{-13}$ eV. Effect of changing lightest neutrino mass on final asymmetry was investigated in earlier works mentioned above, we will discuss this in the context of non-standard cosmology in upcoming sections. To have an overall picture of leptogenesis in standard radiation dominated universe, we perform a numerical scan, the result of which is shown in figure \ref{standardlepto2} which shows that the scale of leptogenesis $M_1$ can be as low as 7 TeV in this case. Similar results can be obtained for inverted ordering of light neutrino mass as well. We will compare the results of leptogenesis in non-standard cosmology in upcoming sections with the plot shown in figure \ref{standardlepto2}.

\section{Early matter dominated universe}
\label{sec:EMD}
\begin{figure}
\begin{center}
\includegraphics[scale=.25]{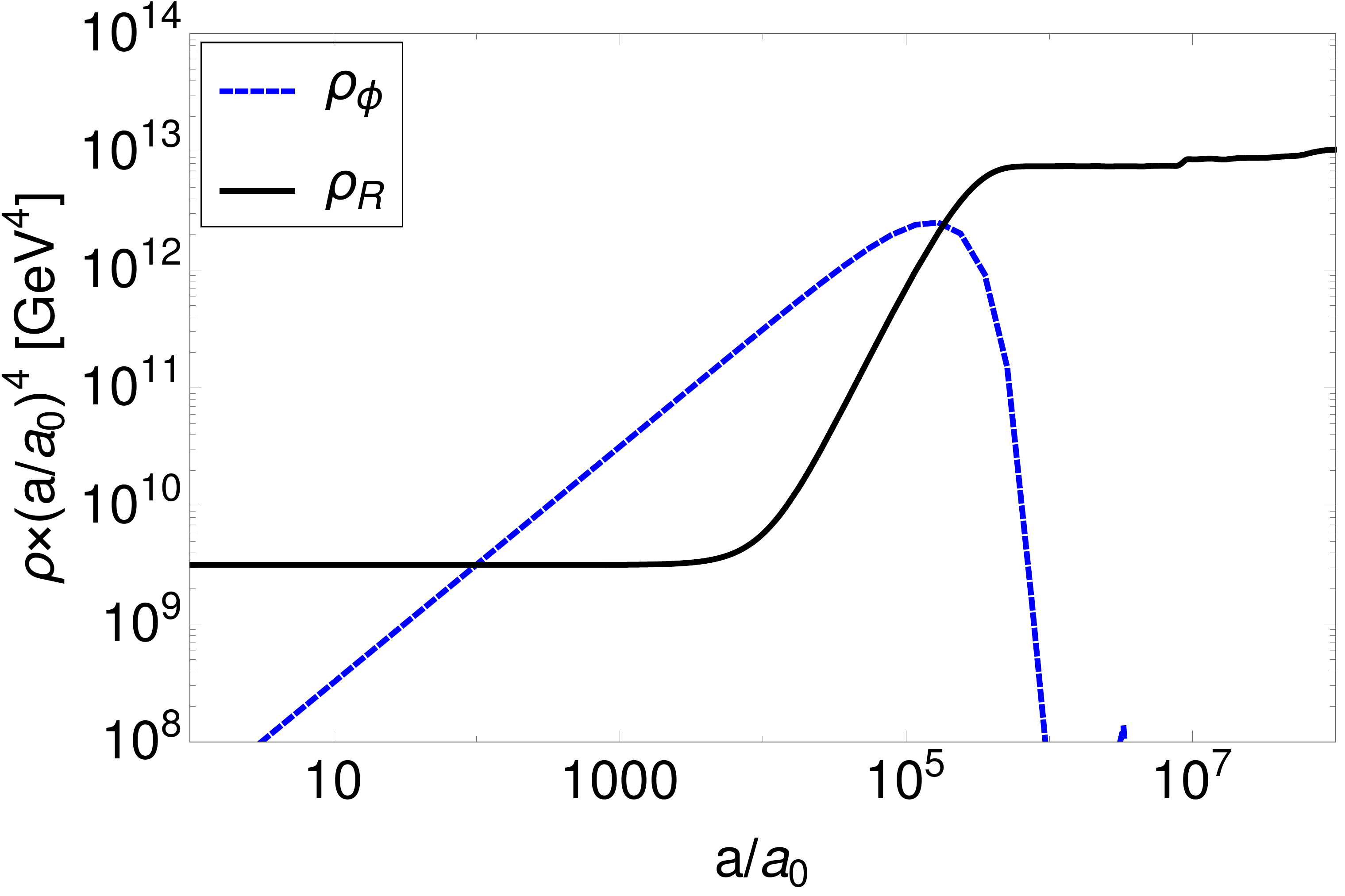}
\includegraphics[scale=.25]{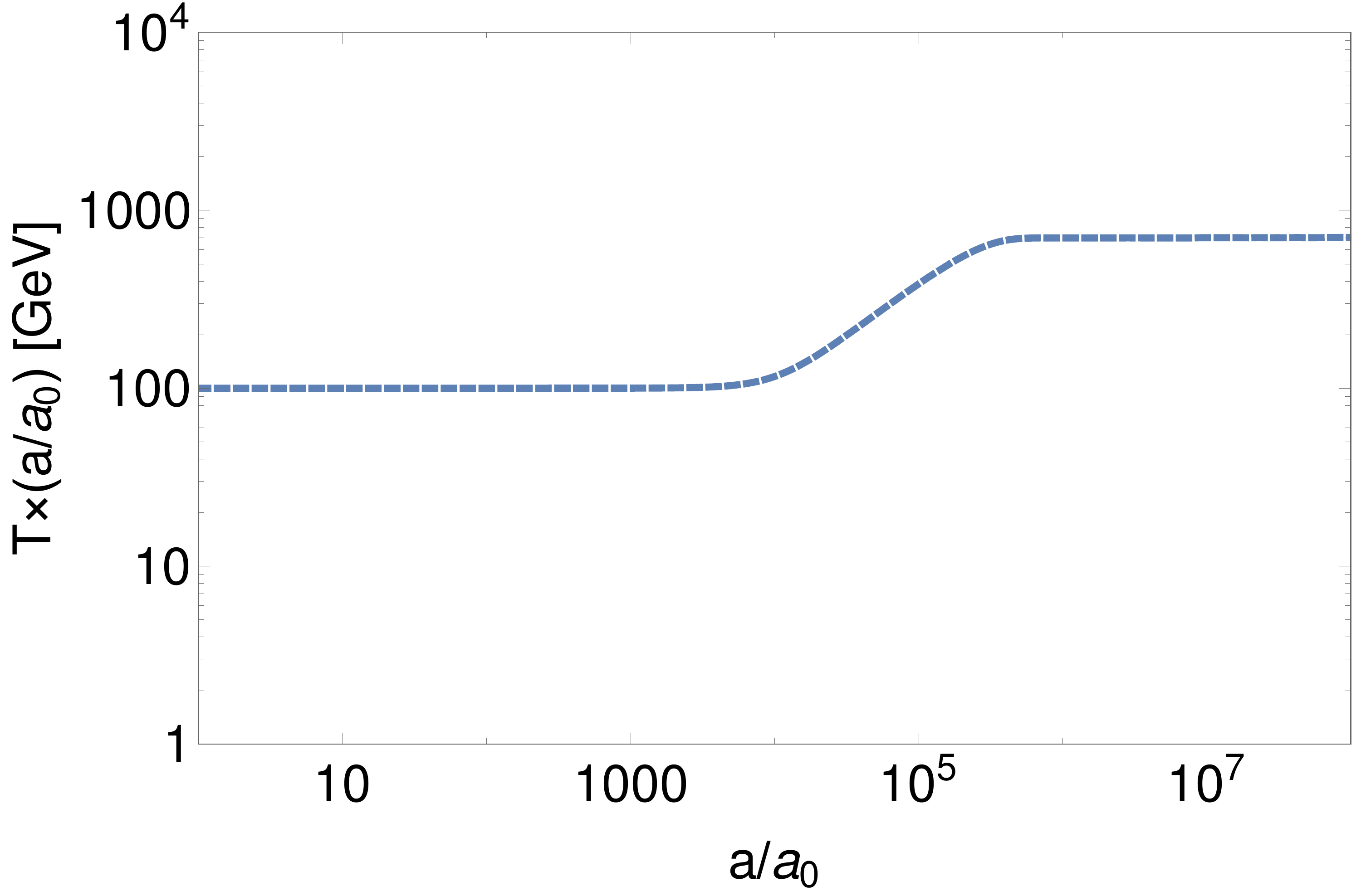}
\caption{Evolution of the energy densities for radiation and the $\phi$ field (left panel), and evolution of radiation temperature T (right panel) as a function of scale factor a, for $\omega=0$, $T_{\rm end}=7\times 10^{-3}$ GeV and $\dfrac{\rho_{\phi}(T=m_{\phi})}{\rho_{R}(T=m_{\phi})}=10^{-2}$.}
\label{emd1a}
\end{center}
\end{figure}

In this scenario, we consider an epoch in the early universe (prior to the BBN era) to be dominated by a matter component instead of a purely radiation dominated era of $\Lambda{\rm CDM}$ cosmology. The total energy density of the universe in this epoch was dominated by a scalar field $\phi$, which behaves like an ordinary pressure-less matter. Known as early matter dominated (EMD) universe, the expansion of the universe in this era is obviously slower compared to the radiation dominated universe of standard cosmology. This is equivalent to the fact that the energy density of the matter field $\rho_{\phi}$ falls with the expansion of the universe at a slower rate compared to the radiation energy density $\rho_{\rm rad}$ as long as $\phi$ does not decay. In principle $\phi$ can decay to both SM radiation and dark sector particles like DM. In the early universe the evolution of $\rho_{\phi}$, the SM entropy density $s$, as well as the DM number density $n$ are governed by the system of coupled Boltzmann equations \cite{Arias:2019uol}

\begin{equation} \label{eq:66a}
\dfrac{d\rho_{\phi}}{dt}+3(1+\omega)H\rho_{\phi}=-\Gamma_{\phi}\rho_{\phi},
\end{equation}

\begin{equation} \label{eq:67a}
\dfrac{ds}{dt}+3Hs= \dfrac{\Gamma_{\phi}\rho_{\phi}}{T} \left( 1-b\dfrac{E}{m_{\phi}} \right) +2\dfrac{E}{T} \langle \sigma v_{\rm rel} \rangle \left( n^{2}-n_{\rm eq}^{2} \right),
\end{equation}

\begin{equation}  \label{eq:68a}
\dfrac{dn}{dt}+3Hn=\dfrac{b}{m_{\phi}}\Gamma_{\phi}\rho_{\phi}- \langle \sigma v_{\rm rel} \rangle (n^{2}-n_{\rm eq}^{2}),
\end{equation}
where $ \langle \sigma v_{\rm rel} \rangle$ is the total DM annihilation cross-section into SM particles and $E^{2}\simeq m^{2}+3T^{2}$ is the averaged energy per DM particle. Here, $b$ is twice the branching ratio $\phi$ decaying into a couple of DM particles and thus $b$ controls the proportion at which $\phi$ decays to SM radiation and DM. $\left( 1-b\dfrac{E}{m_{\phi}} \right)$ is the fraction of $\phi$ energy that goes into radiation. The second term on the RHS of equation \eqref{eq:67a} is the entropy injection due to DM annihilations which is subdominant compared to the first term and hence can be ignored. Also we assume that the branching ratio of $\phi$ decaying to DM particles is very small so that effectively $b=0$. This simplifies the calculation of DM abundance very simple as it is governed by usual annihilation and coannihilation processes, similar to the WIMP paradigm. It also keeps the model minimal. Equation \eqref{eq:67a} plays an important role to track the temperature of the SM plasma through the entropy density $s$
\begin{equation}
s(T)=\dfrac{\rho_{R}+p_{R}}{T}=\dfrac{2\pi^{2}}{30}g_{*s}(T)T^3.
\end{equation}
The evolution of the SM radiation temperature is given by
\begin{equation}
\dfrac{dT}{da}=\left( 1+\dfrac{T}{3g_{*s}}\dfrac{dg_{*s}}{dT} \right)^{-1} \left[ -\dfrac{T}{a}+\dfrac{\Gamma_{\phi}\rho_{\phi}}{3Hsa}\left( 1-\dfrac{Eb}{m_{\phi}} \right)+\dfrac{2}{3}\dfrac{E \langle \sigma v_{\rm rel} \rangle}{Hsa}(n^{2}-n_{\rm eq}^{2}) \right]
\end{equation}
In order not to alter the successful predictions of BBN in standard $\Lambda$CDM cosmology, the temperature at the end of the $\rho_{\phi}$ dominated phase has to be $T_{\rm end}\gtrsim 4$ MeV \cite{Kawasaki:2000en,Hannestad:2004px,Ichikawa:2005vw}, where $T_{\rm end}$ is given by the total decay width $\Gamma_{\phi}$ as
\begin{equation}
T_{\rm end}^{4}=\dfrac{90}{\pi^{2}g
_{*}(T_{\rm end})}M_{\rm Pl}^{2}\Gamma_{\phi}^{2}
\end{equation}
Since we are not defining the specific interactions of $\phi$ with SM particles, its decay width can be kept as a free parameter for a model independent analysis. Therefore, this type of EMD universe can be characterised by two free parameters, $T_{\rm end}$ and $k=\dfrac{\rho_{\phi}(T=m_{\phi})}{\rho_{\rm rad}(T=m_{\phi})}$ as adopted by the authors of \cite{Arias:2019uol} for DM analysis. Since our motivation is to study leptogenesis in such an EMD universe, we define $k=\dfrac{\rho_{\phi}(T=M_{1})}{\rho_{\rm rad}(T=M_{1})}$ so that it is directly related to the scale of leptogenesis. This does not change the DM analysis from \cite{Arias:2019uol} if we assume $b=0$ so that $m_{\phi}$ does not enter the analysis, as mentioned earlier. Before proceeding to analyse different scenarios in the context of leptogenesis and DM in an EMD universe, we show the evolution of radiation and $\phi$ energy densities as well as radiation temperature in figure \ref{emd1a}. Clearly, the energy density of $\phi$ evolves like usual matter until $\phi$ decays completely into radiation giving a sudden increase in radiation energy density. The effect of $\phi$ decay is also visible in the evolution of temperature as radiation temperature increases suddenly due to entropy injection from $\phi$ decay into radiation. The results also matches with the ones shown in \cite{Arias:2019uol}. 

\subsection{Leptogenesis}
Since the entropy per comoving volume of the universe is not constant in this case, we can no longer write the usual Boltzmann equations for leptogenesis. In such a case, the relevant Boltzmann equations are found to be

\begin{equation}\label{eq:a}
\dfrac{dn_{B-L}}{dz}+\dfrac{n_{B-L}}{s}\dfrac{ds}{dz}+\dfrac{3n_{B-L}}{z}=-\epsilon_1 D_1(n_{N_{1}}-n_{N_{1}}^{\rm eq})-W^{\rm Total}n_{B-L}
\end{equation} 

\begin{equation} \label{eq:b}
\dfrac{dn_{N_{1}}}{dz}+\dfrac{n_{N_{1}}}{s}\dfrac{ds}{dz}+\dfrac{3n_{N_{1}}}{z}=-D_1(n_{N_{1}}-n_{N_{1}}^{\rm eq}),
\end{equation}

where 

\begin{equation}
D_1=K_{1}z\dfrac{\kappa_{1}(z)}{\kappa_{2}(z)}\dfrac{H_{\rm rad}(z)}{H(z)}
\end{equation}

\begin{equation}
W^{\rm Total}=W_{\rm ID}+\Delta W
\end{equation}

\begin{equation}
W_{\rm ID}=\dfrac{1}{4}K_{1}z^{3}\kappa_{1}(z)\dfrac{H_{\rm rad}(z)}{H(z)}
\end{equation}

\begin{equation}
\Delta W=\dfrac{36\sqrt{5}M_{\rm Pl}}{\pi^{1/2}g_{l}\sqrt{g_{*}}v^{4}}\dfrac{1}{z^{2}}\dfrac{1}{\lambda_{5}^{2}}M_{1}{\bar{m}}^{2}_{\xi}\dfrac{H_{\rm rad}(z)}{H(z)}
\end{equation}

\begin{equation}
n_{N_{1}}^{2}=\dfrac{z^{2}}{2}\kappa_{1}(z).
\end{equation}

Here $n_{N_{1}}$ and $n_{B-L}$ are the comoving number densities of $N_{1}$ and $B-L$ respectively. The $\kappa_{i}(z)$ are the modified Bessel functions of second kind, same as the ones used earlier. The $W_{\rm ID}$ term in the washout term corresponds to the inverse decay while the $\Delta W$ term corresponds to the $\Delta L=2$ scattering processes.  The equations \eqref{eq:a} and \eqref{eq:b} have to be solved simultaneously with equations \eqref{eq:66a} and \eqref{eq:67a} which we rewrite as 
\begin{equation}\label{eq:71}
z\dfrac{d\rho_{\phi}}{dz}+3(1+\omega)\rho_{\phi}=-\dfrac{\Gamma_{\phi}\rho_{\phi}}{M_1H},
\end{equation}

\begin{equation}\label{eq:72}
z\dfrac{ds}{dz}+3(1+\omega)s=\dfrac{z\Gamma_{\phi}\rho_{\phi}}{M_1H}
\end{equation}
respectively. The Boltzmann equation for DM \eqref{eq:68a} is also solved simultaneously, the results of which will be discussed in upcoming section.

The Hubble parameter, in general, is given by 
\begin{equation}
H(z)=\sqrt{\dfrac{\rho_{\phi}(z)+\rho_{\rm rad}(z)}{3M_{\rm Pl}^{2}}}
\end{equation} 
Though a compact analytical form is difficult to obtain, under the sudden decay approximation it can be written as \cite{Arias:2019uol} 

\[
H(z)=
\begin{cases}
\dfrac{\pi}{3}\sqrt{\dfrac{g_{*}}{10}}\dfrac{M_{1}^{2}}{M_{\rm Pl}}\dfrac{1}{z^{2}} ,              & \text{if }  z \le z_{\rm eq}\\

\dfrac{\pi}{3}\sqrt{\dfrac{g_{*}}{10}}\dfrac{M_{1}^{2}}{M_{\rm Pl}} \sqrt{\dfrac{k}{z^{3(1+\omega)}}},& \text{if } z_{\rm eq}\le z\leq z_{\rm end}\\

    \dfrac{\pi}{3}\sqrt{\dfrac{g_{*}}{10}}\dfrac{M_{1}^{2}}{M_{\rm Pl}}\dfrac{1}{z^{2}} ,              & \text{if }  z \ge z_{\rm end}  .
\end{cases}
\]

Here, $\omega$ is the equation of state parameter for the new species $\phi$, for matter field $\omega=0$ and $k$ is the ratio between $\rho_{\phi}$ and $\rho_{\rm rad}$ at $T=M_{1}$ i.e. at $z=1$. Also, $z_{\rm eq}$ corresponds to the temperature $T_{\rm eq}$ at which the $\rho_{\phi}$ overcomes $\rho_{\rm rad}$, as seen from figure \ref{emd1a}. On the other hand, $z_{\rm end}$ corresponds to $T_{\rm end}$ defined earlier. We however, use the exact Hubble parameter in our numerical analysis. In addition to $T_{\rm end}, T_{\rm eq}$ another temperature relevant for our analysis is the sphaleron freeze-out temperature $T_{\rm sphaleron}$. Depending upon $T_{\rm end}, T_{\rm eq}$, we study three different cases below. Note that, we have not discussed the case where $ T_{\rm end}, T_{\rm eq} \ll T_{\rm sphaleron}$ as this is very similar to leptogenesis in a radiation dominated universe (upto a subsequent entropy dilution). Alternately, if $T_{\rm end}, T_{\rm eq}$ are much larger than the scale of leptogenesis $T=M_1$, then also it resembles the usual scenario already studied in several earlier works.
\begin{figure}[h]
\begin{center}
\includegraphics[scale=.23]{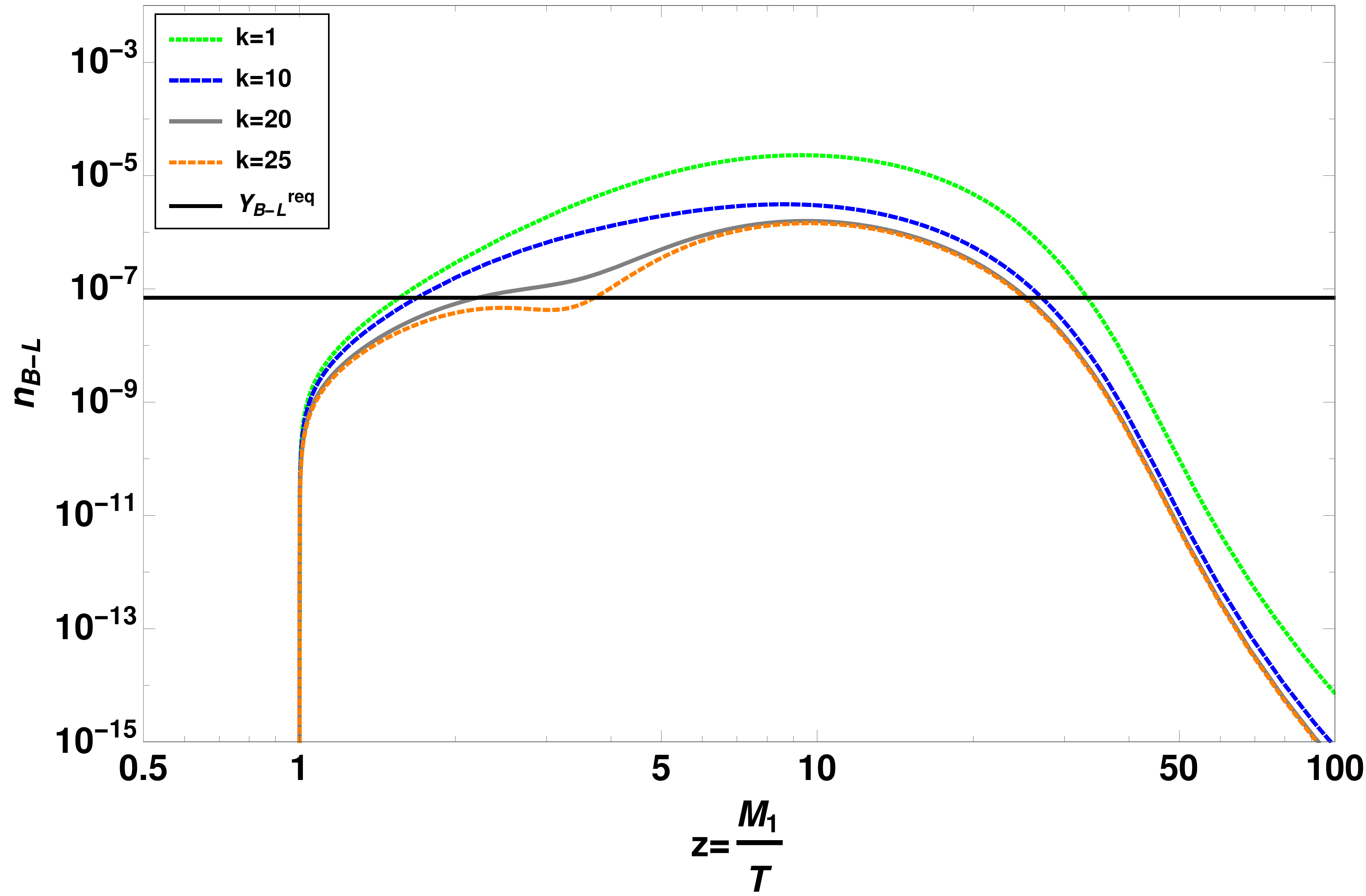}
\includegraphics[scale=.23]{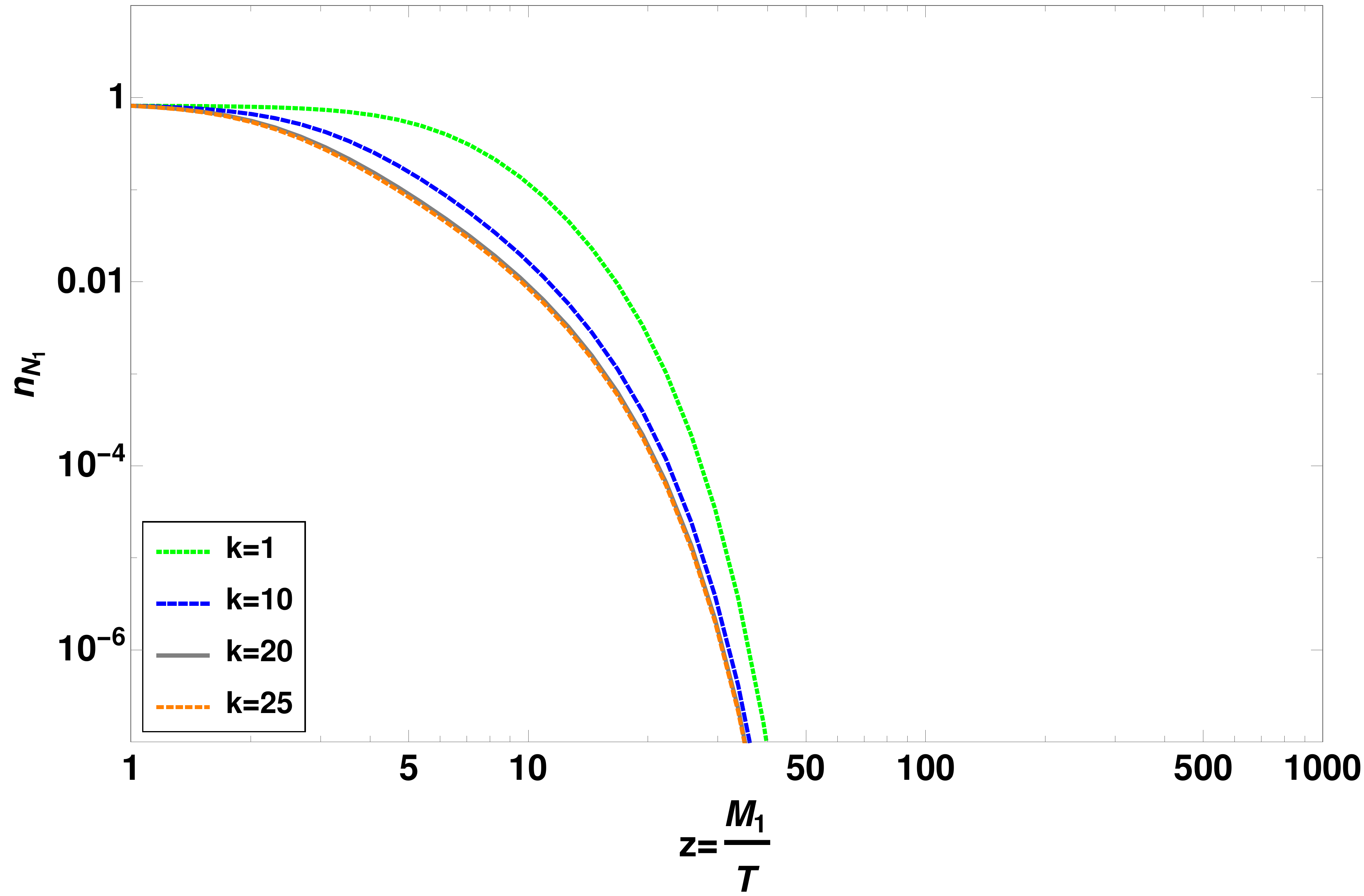}
\caption{Evolution of $n_{B-L}$ (left panel) and $n_{N_{1}}$ (right panel) with $z=\dfrac{M_{1}}{T}$ in EMD universe (Case 1). The parameters used for these plots are $m_{1}=10^{-13}$ eV, $M_{1}=2\times10^{4}$ GeV, $M_{i}/M_{i+1}=10^{0.5}$, $\lambda_{5}=10^{-4}$ and $T_{\rm end}=10^{3}$ GeV.}
\label{fig:1}
\end{center}
\end{figure}
\begin{figure}[h]
\begin{center}
\includegraphics[scale=.23]{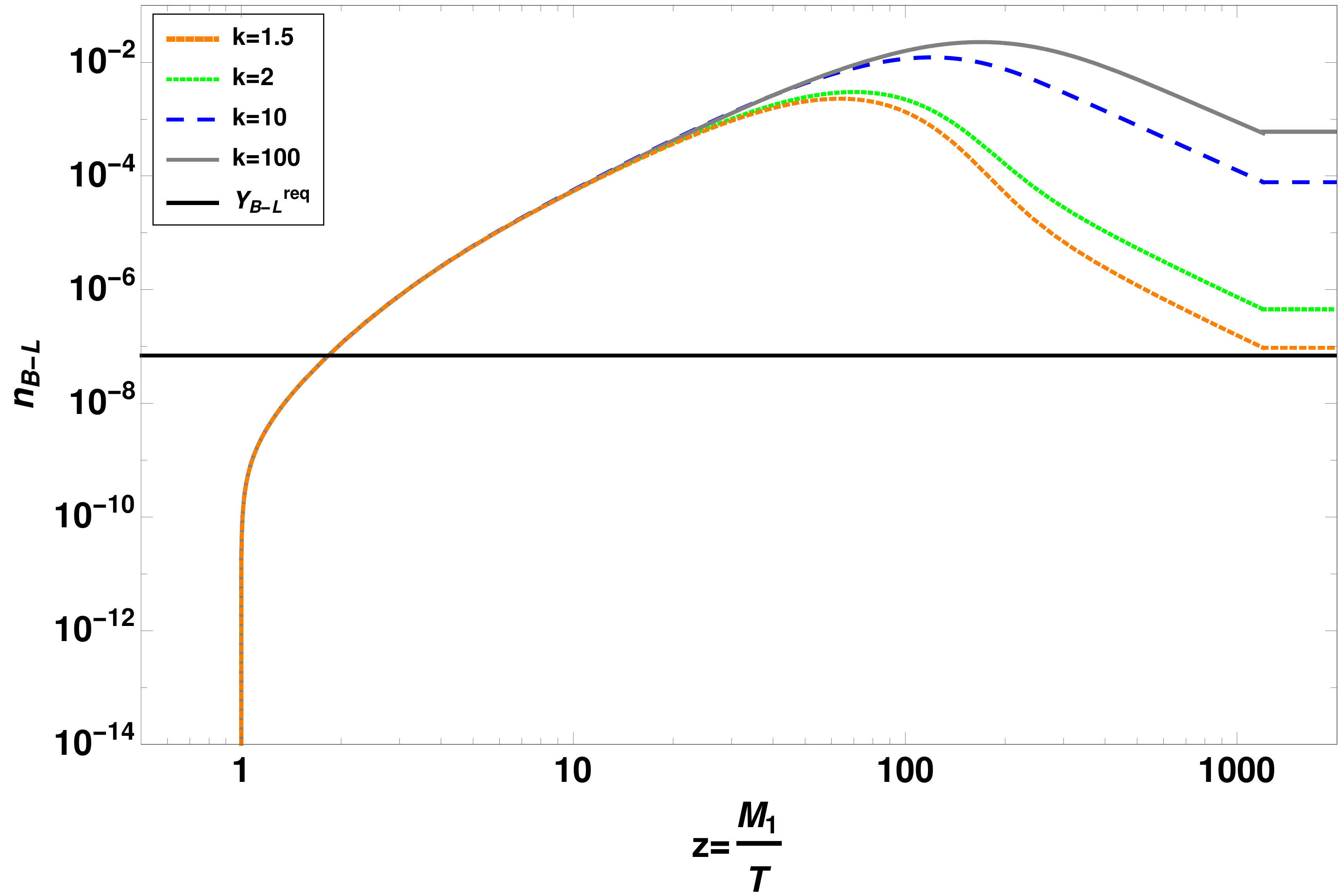}
\includegraphics[scale=.23]{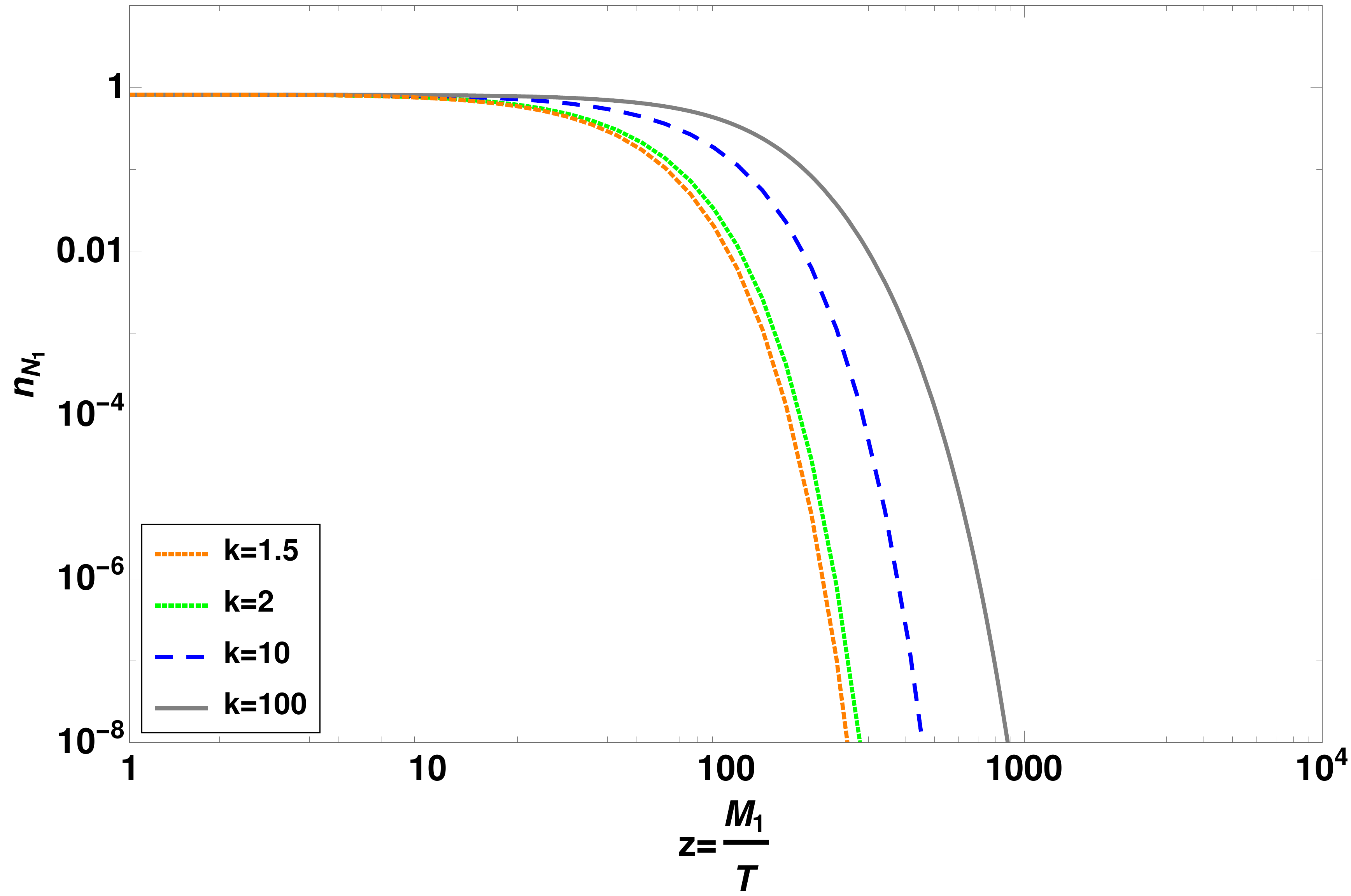}
\caption{Evolution of $n_{B-L}$ (left panel) and $n_{N_{1}}$ (right panel) with $z=\dfrac{M_{1}}{T}$ in EMD universe (Case 2). The parameters used for these plots are $m_{1}=10^{-13}$ eV, $M_{1}=3\times10^{5}$ GeV, $M_{i}/M_{i+1}=10^{0.5}$, $\mu_2=3500$ GeV, $\lambda_{5}=10^{-4}$ and $T_{\rm end}=250$ GeV.}
\label{fig:2emd}
\end{center}
\end{figure}

\begin{figure}[h]
\begin{center}
\includegraphics[scale=.30]{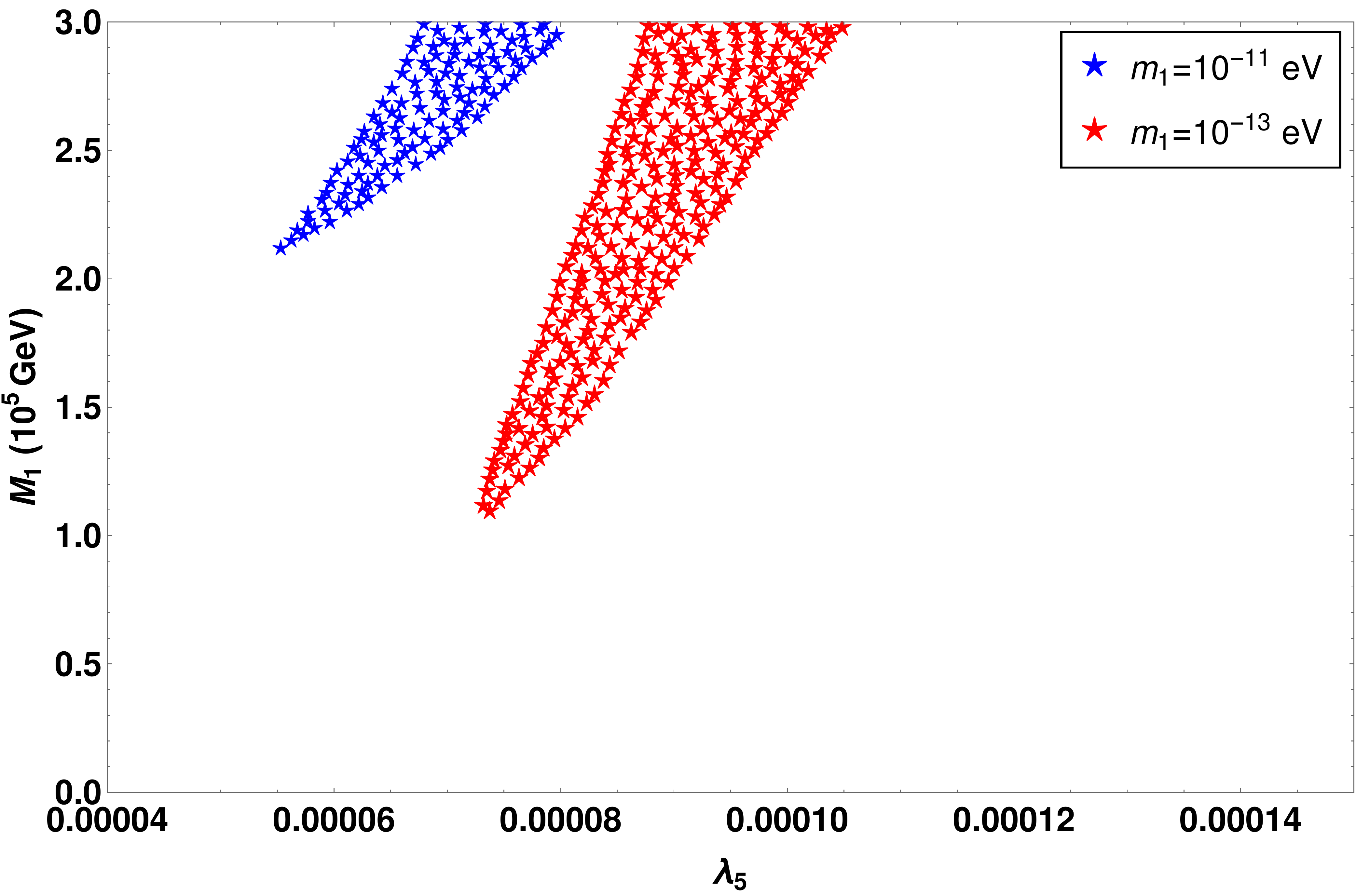}

\caption{Parameter space in $M_{1}-\lambda_{5}$ plane giving rise to the observed baryon asymmetry in EMD universe (Case 2). The benchmark parameters used for this result are $\mu_2=3500$ GeV and $M_{i+1}/M_{i}=10^{0.5}$. The cosmological parameters used for these results are $k=1.7$ and $T_{\rm end}=250$ GeV.}
\label{fig:2emda}
\end{center}
\end{figure}

\begin{figure}[h]
\includegraphics[scale=.4]{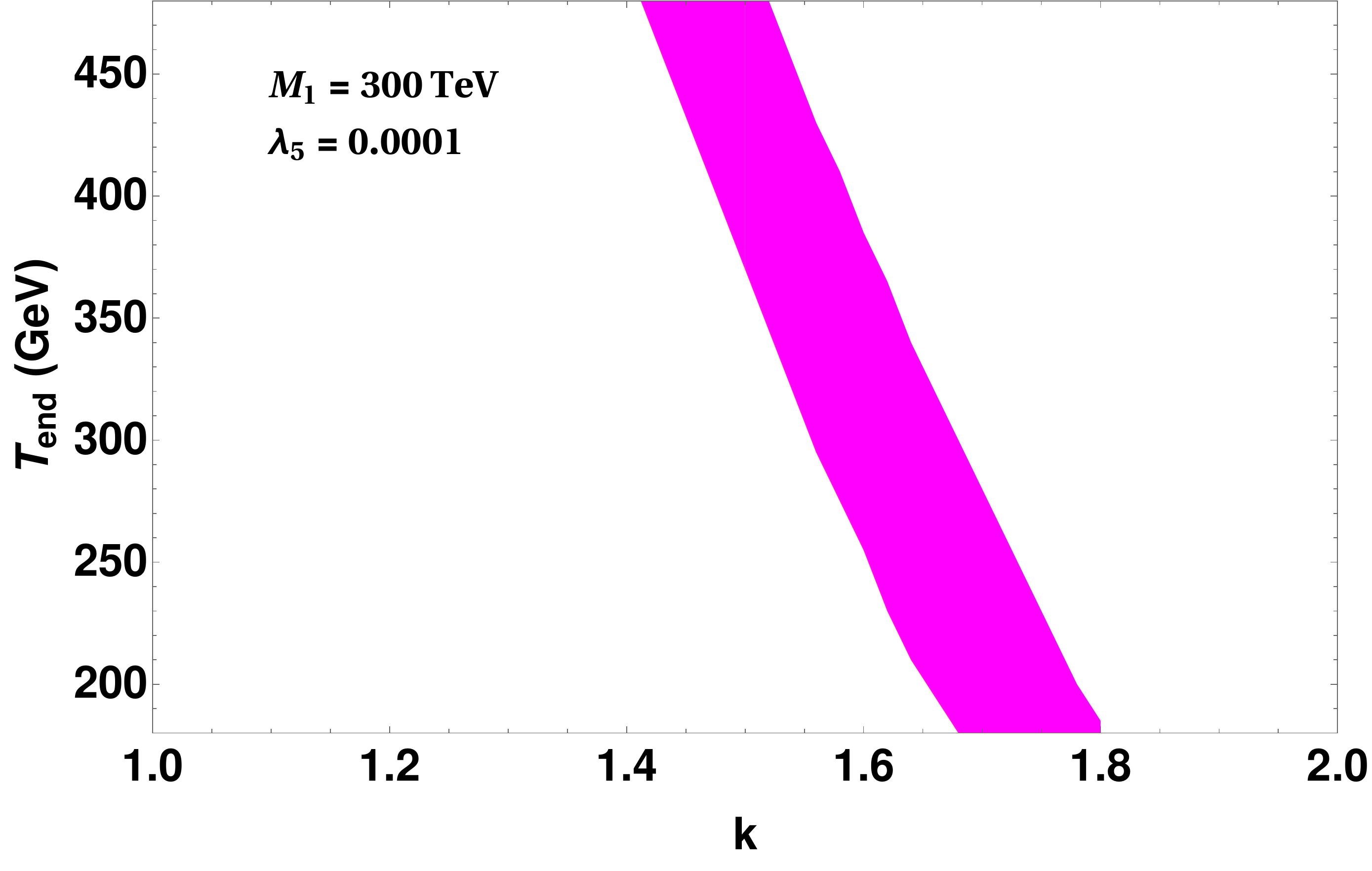}
\caption{Allowed parameter space in $k-T_{\rm end}$ plane for successful leptogenesis in EMD universe (Case 2). The benchmark parameters used for this result are $M_{1}=300$ TeV, $M_{i+1}/M_{i}=10^{0.5}$, $\lambda_5=0.0001$, $\mu_{2}=3500$ GeV and $m_{1}=10^{-13}$ eV. }
\label{fig:2emdb}
\end{figure}

\subsubsection{Case 1: $T_{\rm sphaleron} \ll T_{\rm end} \ll T_{\rm eq}$}

Here, we have taken $T_{\rm end}=10^{3}$ GeV with $M_{1}=20$ TeV, that is $z_{\rm end}=20$. However, the $z_{\rm sphaleron}\simeq 100$. In this case the expansion of the universe is mainly determined by $\rho_{\phi}$ till $z_{\rm end}=20$ and then it is mainly determined by the usual radiation. One can observe from the evolution plots shown in figure \ref{fig:1} that the generated $B-L$ and also $n_{N_{1}}$ start undergoing a sharp dilution from around $z\sim20$ because of the entropy injection from $\phi$ decay. We can see that larger values of $k$ lead to more dilution, which is expected as larger relative abundance of $\phi$ over radiation will inject more entropy to the SM radiation when it decays. Even if we increase $M_1$ to a higher value like $10^{12}$ GeV, we do not get the required asymmetry due to large entropy dilution around $z_{\rm end}$. Since low scale leptogenesis is not possible in this case, we do not discuss it further from both leptogenesis as well as DM point of view.

\subsubsection{Case 2: $T_{\rm sphaleron}\lesssim T_{\rm end} \ll T_{\rm eq}$}
In the previous case, we found that keeping $T_{\rm end}$ much above $T_{\rm sphaleron}$ leads to significant dilution of lepton asymmetry resulting in negligible baryon asymmetry after sphaleron transitions. Therefore, we now choose $T_{\rm end}$ to be closer to $T_{\rm sphaleron}$ so that not all the generated lepton asymmetry get significantly diluted before sphaleron freeze-out. In figure \ref{fig:2emd}, we show the corresponding evolution of lepton asymmetry as well as $N_1$ abundance by choosing $T_{\rm end}=250$ GeV with $M_{1}=3 \times 10^5$ GeV so that $z_{\rm end}\simeq 1200$. One can observe from the evolution plots that the generated $B-L$ start undergoing a dilution at around $z\sim 100$ because of the entropy injection from $\phi$ decay as well as washout effects. Since the $z_{\rm sphaleron}\simeq 2000$, we have shown the evolution upto $z=2000$. Since the EMD phase disappears at $z_{\rm end}$, there is no further dilution of asymmetry beyond this temperature with the washout effects also getting inefficient. Keeping $k, T_{\rm end}$ fixed at $1.7$, $250$ GeV respectively, we perform a numerical scan for two different values of lightest neutrino mass and show the resulting parameter space in $M_1-\lambda_5$ plane in figure \ref{fig:2emda}. Comparing with the standard radiation dominated scenario discussed earlier, it is clear that the scale of leptogenesis is pushed up by an order of magnitude in this case.

From the evolution plot shown in figure \ref{fig:2emd}, it is observed that the asymmetry is overproduced for higher values of $k$. While the corresponding entropy dilution is also larger for larger values of $k$, yet it is not effective enough to bring it down to the observed limits. To understand the dependence of lepton asymmetry with $k$ as well as $T_{\rm end}$, we vary them by fixing the model parameters at benchmark values. The corresponding result is shown in figure \ref{fig:2emdb} which indicates that large values of $T_{\rm end}$ requires smaller $k$ values to satisfy the correct baryon asymmetry. This seems to be obvious as entropy injection effects are dominant for higher $T_{\rm end}$ (as long as it is below the scale of leptogenesis) and large $k$ values inject more entropy into the plasma.
\begin{figure}[h]

\begin{center}

\includegraphics[scale=.23]{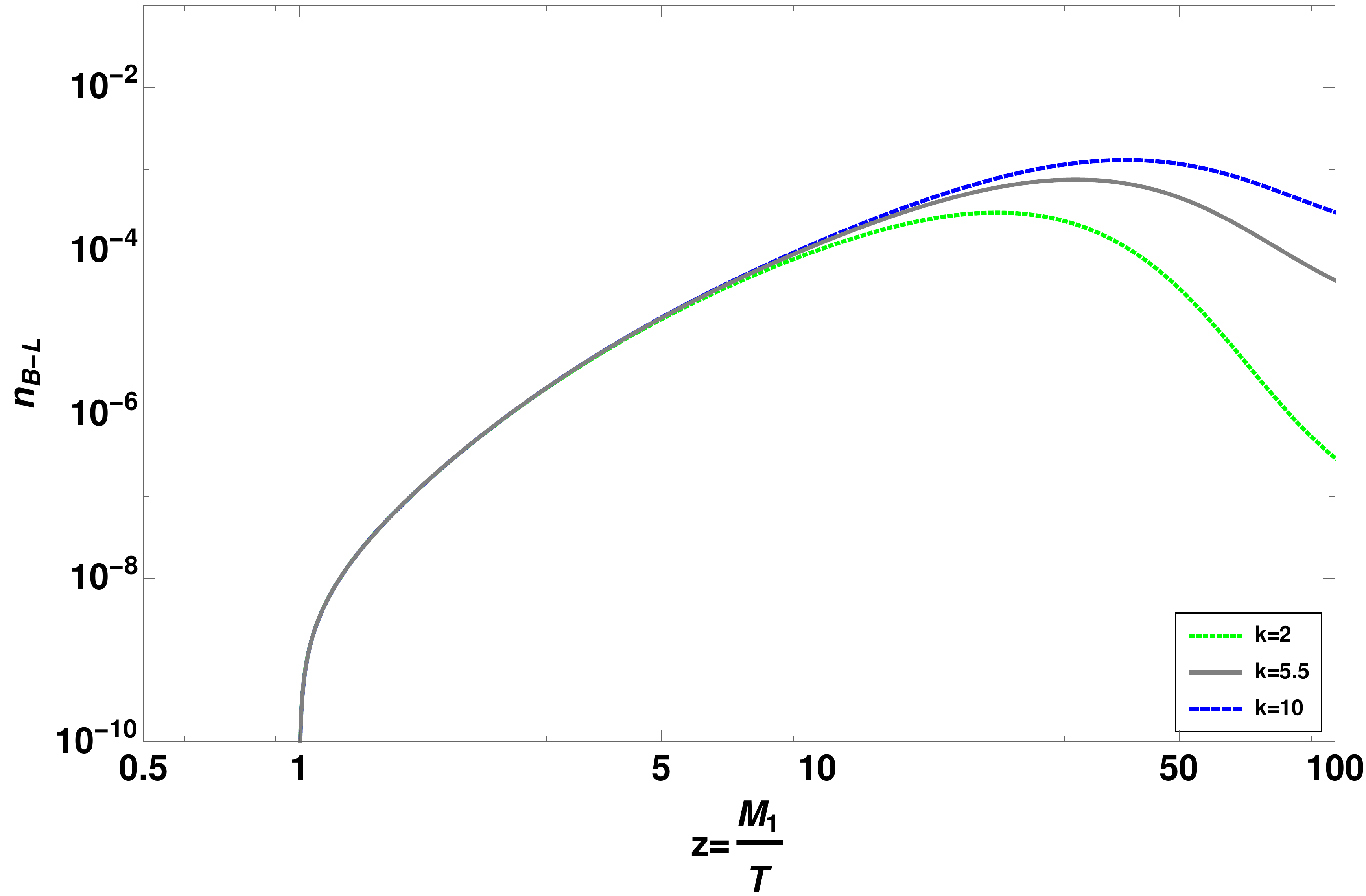}
\includegraphics[scale=.23]{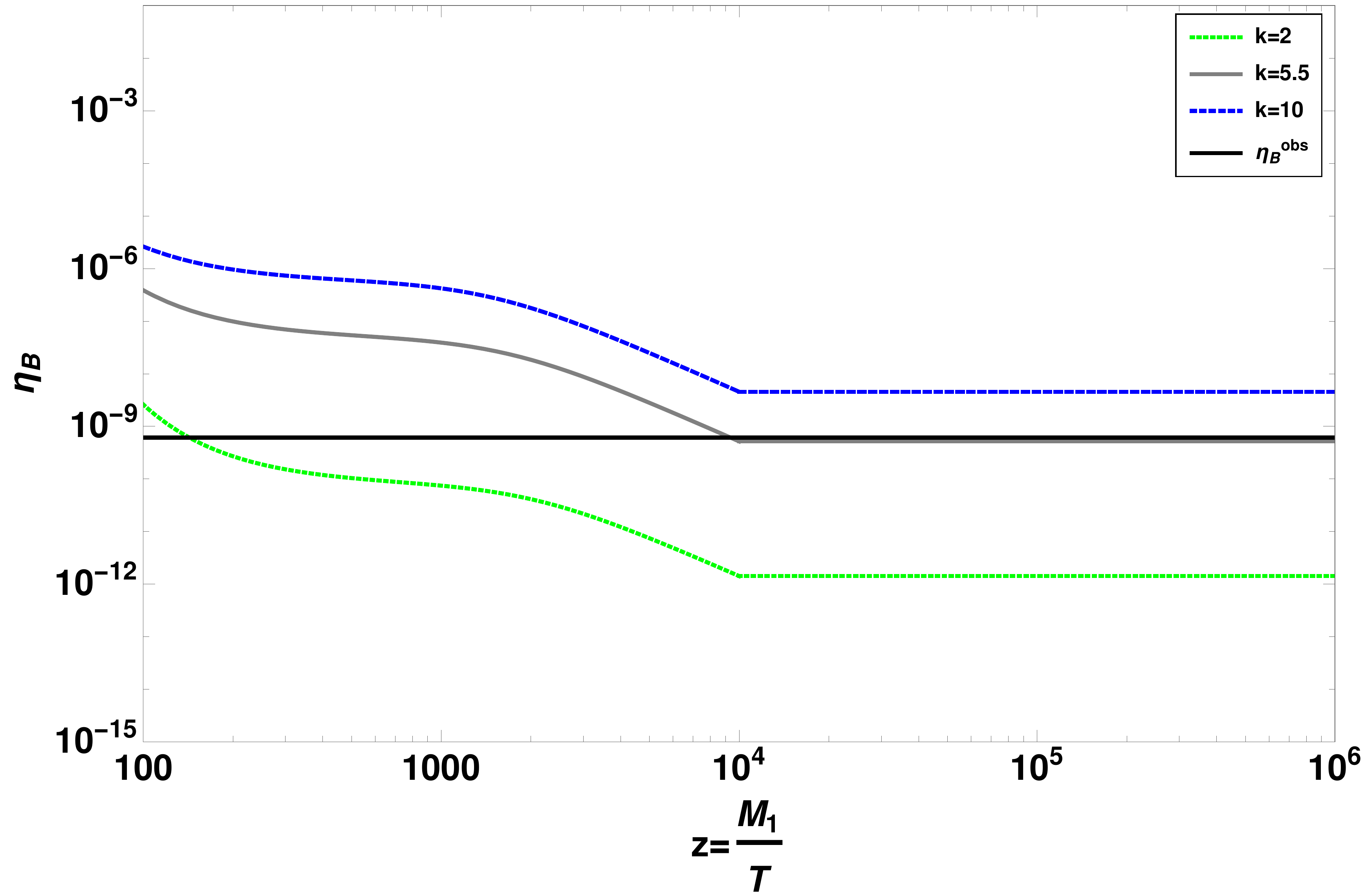}\\
\includegraphics[scale=.23]{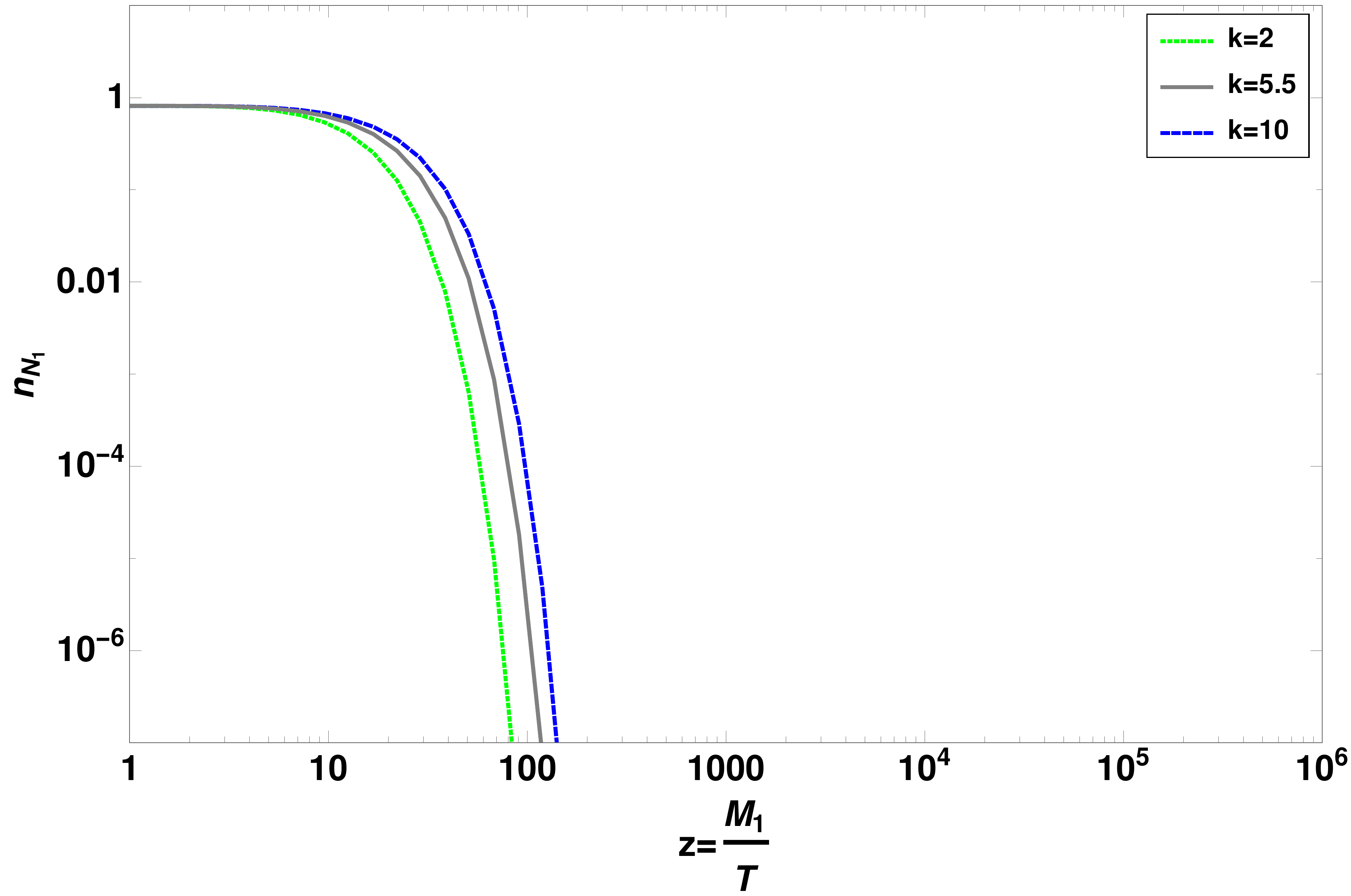}

\caption{Evolution of $n_{B-L}$ (left panel) and $n_{N_{1}}$ (right panel) with $z=\dfrac{M_{1}}{T}$ in EMD universe (Case 3). The parameters used for these plots are $m_{1}=10^{-13}$ eV, $M_{1}=2\times10^{4}$ GeV, $M_{i}/M_{i+1}=10^{0.5}$, $\mu_2=1000$ GeV, $\lambda_{5}=3.5\times10^{-5}$ and $T_{\rm end}=1$ GeV.}
\label{fig:3emd}
\end{center}
\end{figure} 

\begin{figure}
\begin{center}
\includegraphics[scale=.30]{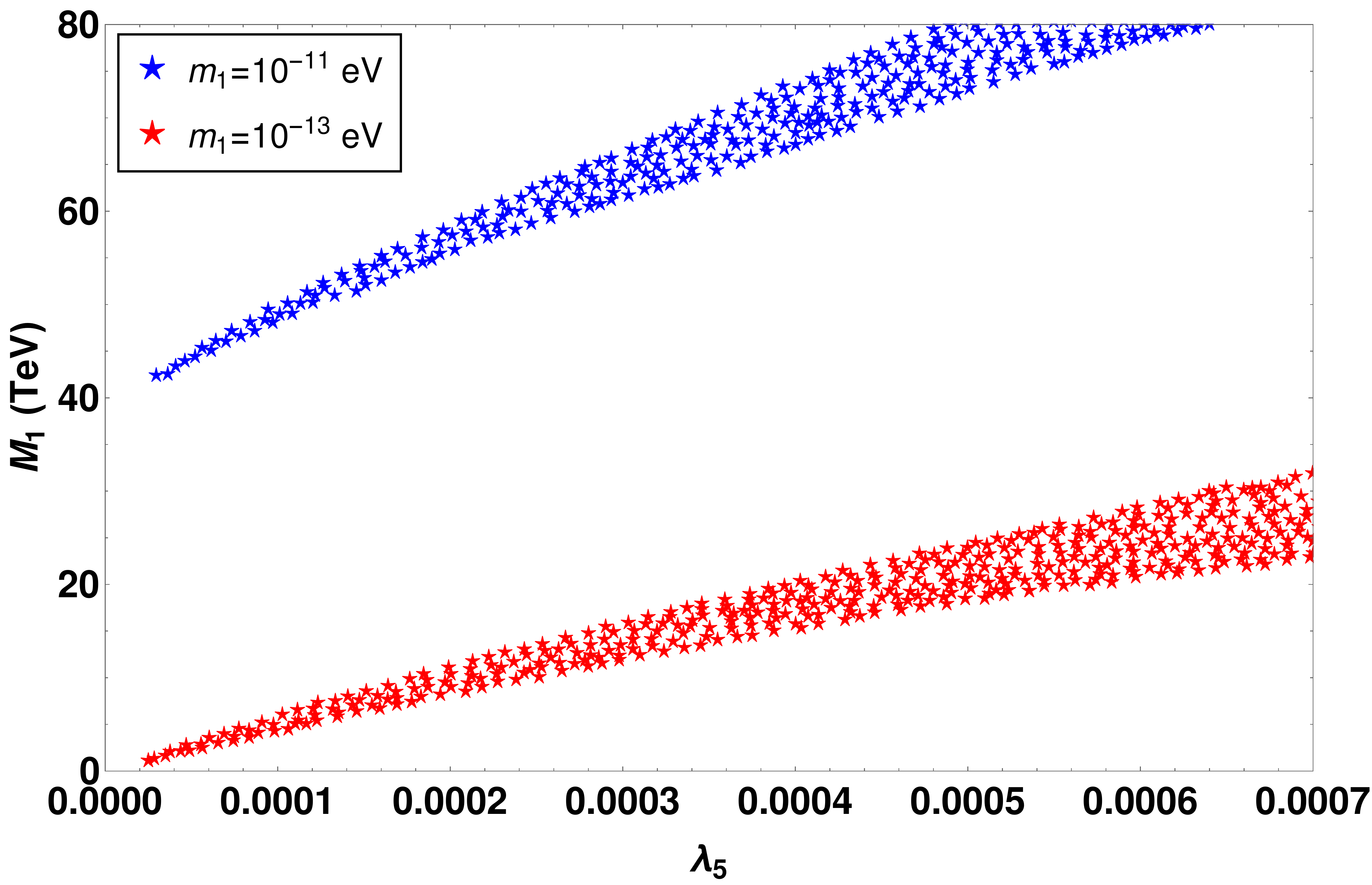}

\caption{Parameter space in $M_{1}-\lambda_{5}$ plane giving rise to the observed baryon asymmetry in EMD universe (Case 3). The benchmark parameters used for these result is $M_{i+1}/M_{i}=10^{0.5}$ and $\mu_2=1000$ GeV. The two cosmological parameters used for this result are $k=5.5$ and $T_{\rm end}=1$ GeV.}
\label{fig:3emda}
\end{center}
\end{figure}

\begin{figure}[h]
\begin{center}
\includegraphics[scale=.35]{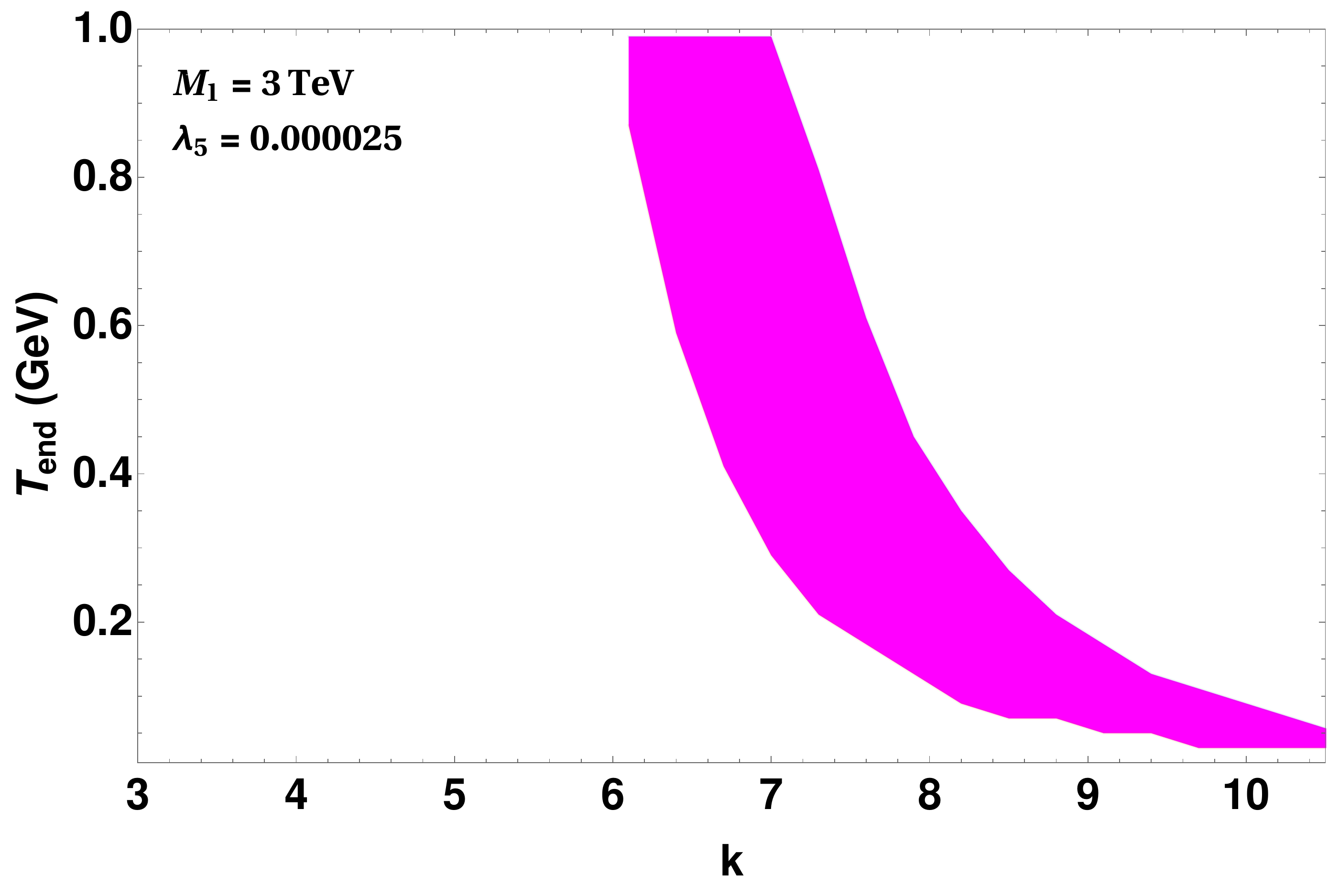}
\caption{Allowed parameter space in $T_{\rm end}-k$ plane for the parameters $m_{1}=10^{-13}$ eV and $\mu_{2}=1000$ GeV in EMD universe (Case 3).}
\label{fig:3emdb}
\end{center}
\end{figure}

\subsubsection{Case 3: $T_{\rm end} \ll T_{\rm sphaleron} \ll T_{\rm eq}$}
We further lower $T_{\rm end}$ and consider the scenario where it can be even lower than $T_{\rm sphaleron}$.
For numerical analysis, we choose $T_{\rm end}=1$ GeV with $M_{1}=20$ TeV, which corresponds to $z_{\rm end}=2\times10^{4}$. Here $z_{\rm sphaleron}\sim 100$ and hence we show the evolution of $n_{B-L}$ upto $z_{\rm sphaleron}$ (top left panel of figure \ref{fig:3emd}) and then $\eta_{B}$ from $z=10^2$ to $z=10^{6}$ (top right panel of figure \ref{fig:3emd}). Since $T_{\rm end}$ is much below the sphaleron freeze-out temperature, there is not much dilution of lepton asymmetry due to entropy injection. The decrease of lepton asymmetry in top left panel plot of figure \ref{fig:3emd} is primarily due to washout effects. The entropy dilution is more visible in the baryon asymmetry shown in top right panel plot of figure \ref{fig:3emd}. Interestingly, even after such dilution, one can have observed baryon asymmetry for $k \approx 5$. The corresponding evolution of $N_1$ number density is shown in bottom panel plot of figure \ref{fig:3emd}. Keeping $k, T_{\rm end}$ fixed at $5$, $1$ GeV respectively, we perform a numerical scan for two different values of lightest neutrino mass and show the resulting parameter space in $M_1-\lambda_5$ plane in figure \ref{fig:3emda}. Comparing with the standard radiation dominated scenario discussed earlier, it is clear that the scale of leptogenesis can be lowered down to around 1 TeV in this case. We then constrain the cosmological parameters in $T_{\rm end}-k$ plane by fixing the model parameters at benchmark values. The corresponding result is shown in figure \ref{fig:3emdb}. Similar to case 2, here also large values of $T_{\rm end}$ requires smaller $k$ values to satisfy the correct baryon asymmetry. 
 \begin{figure}[h]
\begin{center}
\includegraphics[scale=.35]{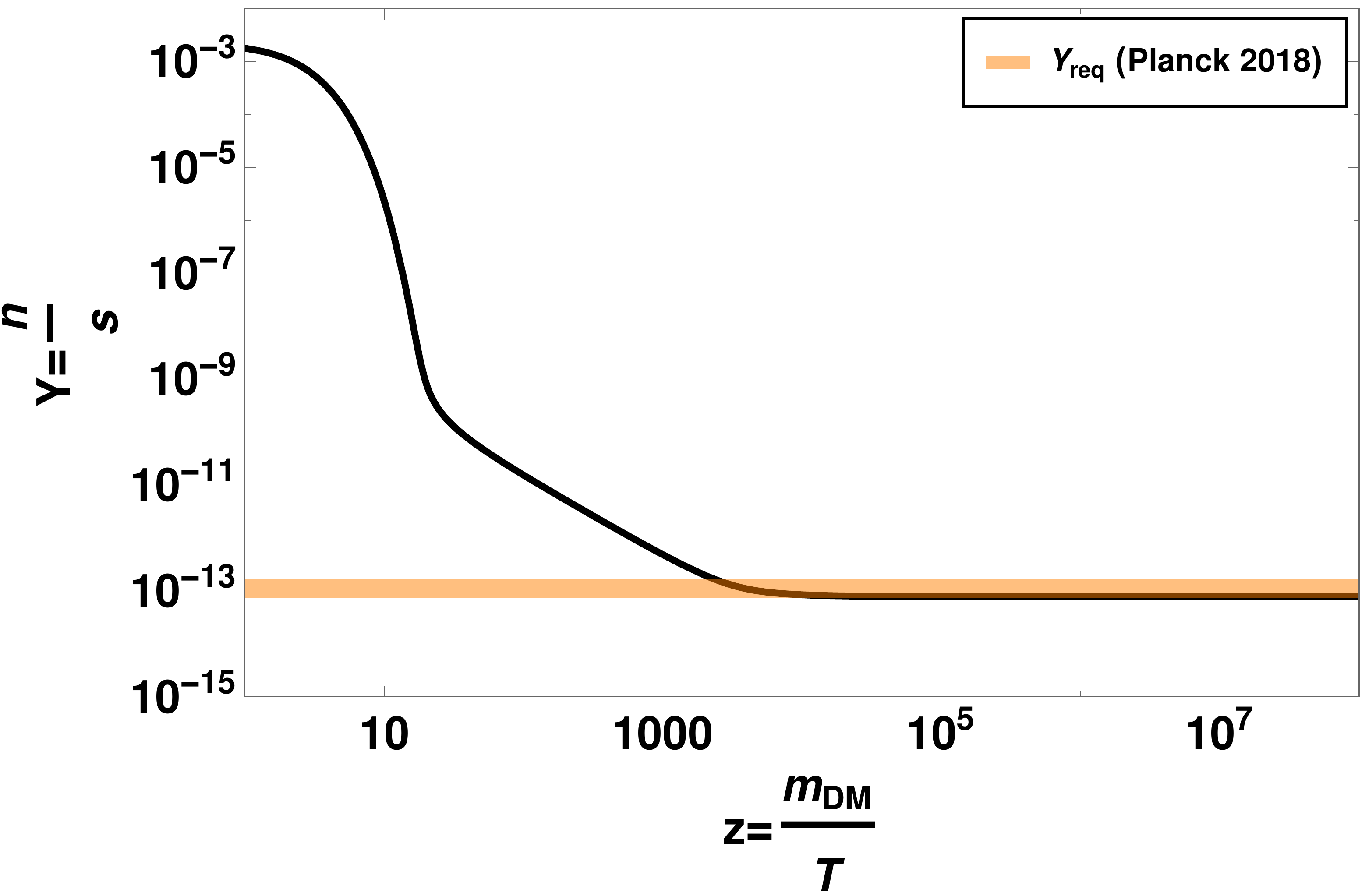}
\caption{Comoving number density of dark matter with $z=m_{\rm DM}/T$ in EMD universe (case 2). The cosmological parameters used for this results are $T_{\rm end}= 250$ GeV and $k=1.7$. The particle physics parameters chosen for this plot are $m_{\rm DM}=3500$ GeV, $m_{H_{0}}=3500.001$ GeV ($\lambda_5=0.0001$), $m_{H^{\pm}}=3500.002$ GeV, $\lambda_{L}=10^{-8}$ and $\lambda_{2}=10^{-2}$. The horizontal band corresponds to Planck 2018 limit on DM abundance \cite{Aghanim:2018eyx}.}
\label{figdm:emd2}
\end{center}
\end{figure}

\begin{figure}[h]
\begin{center}
\includegraphics[scale=.35]{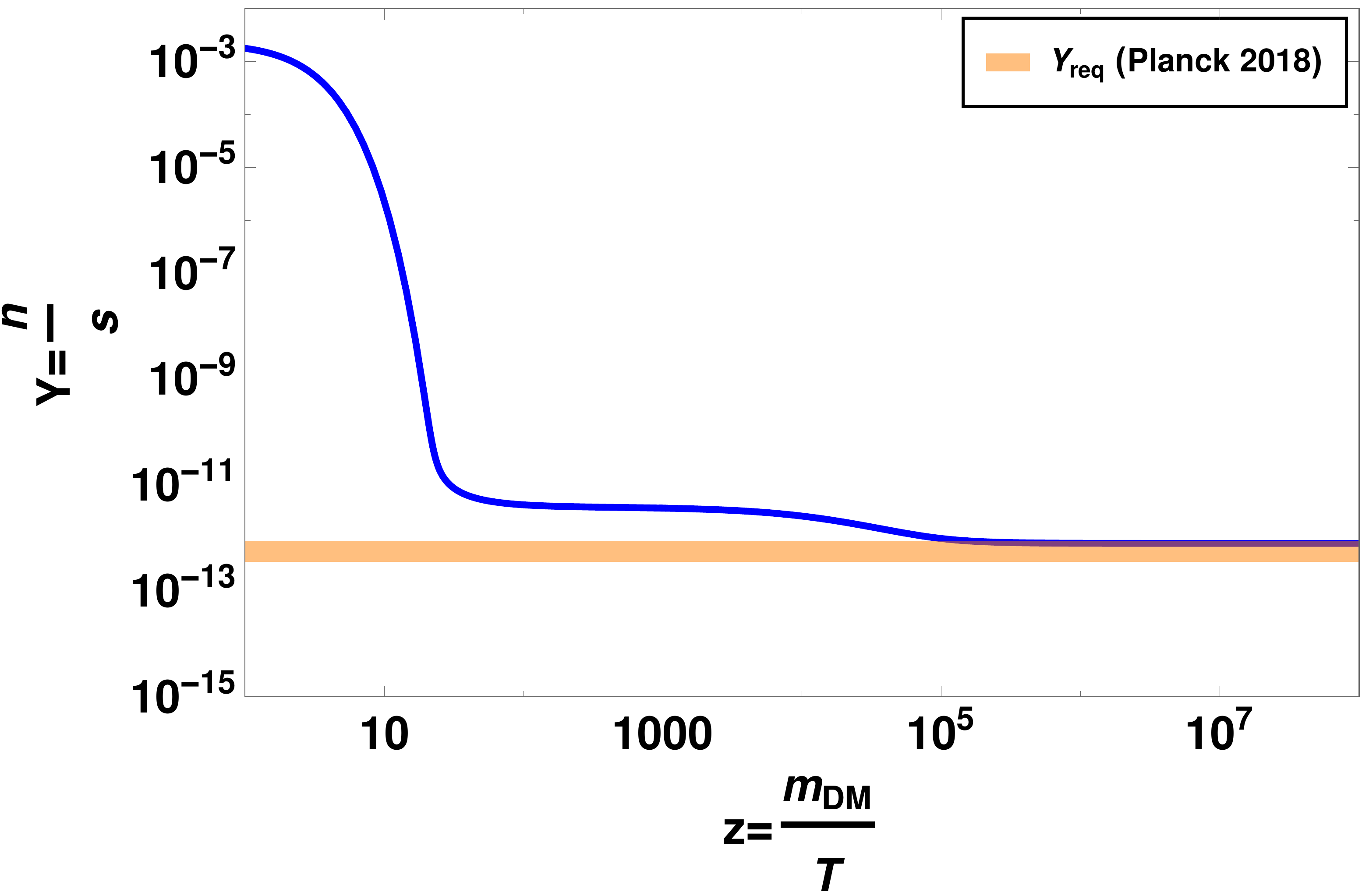}
\caption{Comoving number density of dark matter with $z=m_{\rm DM}/T$ in EMD universe (case 3). The cosmological parameters used for this results are $T_{\rm end}= 1$ GeV and $k=5.5$. The particle physics parameters chosen for this plot are $m_{\rm DM}=1000$ GeV, $m_{H_{0}}=1000.01$ GeV ($\lambda_5=0.0002$), $m_{H^{\pm}}=1000.02$ GeV $, \lambda_{L}=10^{-8}$ and $\lambda_{2}=10^{-2}$. The horizontal band corresponds to Planck 2018 limit on DM abundance \cite{Aghanim:2018eyx}.}
\label{figdm:emd3}
\end{center}
\end{figure}

\subsection{Dark Matter in EMD universe}
Although calculation of DM relic abundance in EMD universe has already been done, here we calculate it for the DM candidate specific to our model and corresponding to the scenarios for leptogenesis discussed above. The relic abundance of DM can be calculated by solving the corresponding Boltzmann equation for comoving number density $Y=n/s$
\begin{equation}\label{eq:70}
\dfrac{dY}{dz}=-\dfrac{\left\langle \sigma v_{\rm rel} \right\rangle s}{Hz}\left( Y^{2}-Y^{2}_{\rm eq} \right).
\end{equation}
where $z=m_{\rm DM}/T$. The Hubble parameter, as defined earlier, is
\begin{equation}
H=\sqrt{\dfrac{\rho_{\phi}+\rho_{\rm rad}}{3M_{\rm Pl}^{2}}}.
\end{equation}
Similar to leptogenesis we solved equations \eqref{eq:70},\eqref{eq:71} and \eqref{eq:72} simultaneously to calculate the abundance of dark matter for benchmark values of DM mass and parameters. The annihilation cross sections of DM $\langle \sigma v_{\rm rel} \rangle$ are evaluated using \texttt{micrOMEGAs} package \cite{Belanger:2013oya}, as mentioned before. We now consider three different scenarios adopted for leptogenesis earlier. Since the scenario with $T_{\rm sphaleron} \ll T_{\rm end} \ll T_{\rm eq}$ (case 1) does not give rise to low scale leptogenesis, we do not study DM details in this work.

For case 2: $T_{\rm sphaleron}\lesssim T_{\rm end} \ll T_{\rm eq}$, we choose $T_{\rm end}=250$ GeV and $k=1.7$ as before. The mass of the DM is chosen to be $m_{\rm DM}=3500$ GeV. The benchmark values are chosen in a way which also gives correct lepton asymmetry. The evolution of DM number density is shown in figure \ref{figdm:emd2}. Similarly for case 3: $T_{\rm end} \ll T_{\rm sphaleron} \ll T_{\rm eq}$, by choosing $T_{\rm end}=1$ GeV and $k=5$ as before, we show the evolution of DM density in figure \ref{figdm:emd3} for $m_{\rm DM}=1000$ GeV. From both these figures, it is clear that the usual freeze-out relic is diluted due to the entropy injection. The choice of benchmark for DM mass is such that its freeze-out occurs at a temperature close to $T_{\rm end}$ so that the entropy dilution effects are clearly visible. Also, the benchmark values of DM parameters are chosen in such a way that the usual thermal relic is overproduced so that even after entropy dilution, the observed DM relic is generated. It is well known for inert doublet DM that in the high mass regime, DM relic is usually overproduced for small mass splitting or $\lambda_5$ \cite{Ma:2006km, Dasgupta:2014hha, Cirelli:2005uq, Barbieri:2006dq, Ma:2006fn,  LopezHonorez:2006gr,  Hambye:2009pw, Dolle:2009fn, Honorez:2010re, LopezHonorez:2010tb, Gustafsson:2012aj, Goudelis:2013uca, Arhrib:2013ela, Diaz:2015pyv, Ahriche:2017iar}. As found earlier, such small values of $\lambda_5$ is also preferred by low scale leptogenesis requirements.

\section{A fast expanding universe}
\label{sec:feu}
%There has been some considerable amount of work in the last few decades for DM searches, but studies have yielded no overwhelming results for what DM actually is. In the previous decades the scenerio for DM which by far got the biggest attention is the WIMP scenerio. In this scenario the DM particles have interaction and masses in the electroweak scale. The correct relic can be obtained if $<\sigma v_{rel}>\simeq 3\times 10^{-26} cm^{3}sec^{-1}$ .The worldwide program for detecting WIMP DM using a multi-channel and multi-messenger approach has followed three main strategies: direct detection, indirect detection, and production at colliders. \\

%However, the observed DM abundance may have generated by decays of some heavy particle by so called freeze-in mechanism. Another simple way to evade the experimental constraints on DM is to consider non-standard cosmological histories, for example scenarios where the Universe was effectively not radiation dominated at an early stage. There are no reasons to assume that the Universe was radiation-dominated prior to Big Bang Nucleosynthesis (BBN). In fact, production of DM in scenarios with non-standard expansion phase of the universe has recently gained increasing interest. Apart from the dark matter, we are going to investigate the effects of non-standard cosmology on leptogenesis in the scotogenic model. 

%\subsection{A faster expansion of the universe}
The other non-standard cosmological epoch we study in this work is the one where prior to the BBN era that is typically around 1 s after the big bang, the universe was dominated by some scalar field $\phi$ instead of usual radiation such that the energy density red-shifts with the scale factor $a$ as follows
\begin{equation}
\rho_{\phi}\propto a^{-(4+n)}.
\label{feu1eq}
\end{equation}
In the above expression, $n>0$. Such a possibility (coined as fast expanding universe) where the energy density at early epochs redshifts faster than radiation leading to $\phi$ domination at early universe but negligible at later epochs was first discussed in the context of WIMP dark matter by the authors of \cite{DEramo:2017gpl}. This is also extended to non-thermal or freeze-in DM models in \cite{DEramo:2017ecx}. In the above expression, $n=0$ corresponds to a universe similar to the usual radiation dominated universe.

The expansion rate of the universe, quantified by the Hubble parameter $H$, is controlled by the total energy density through the Friedmann equations. In FEU scenario where two different species populate the early universe, the total energy density in the very early epochs can be written as
\begin{equation}
\rho(T)=\rho_{\rm rad}(T)+\rho_{\phi}(T)
\end{equation}
where the usual radiation energy density $\rho_{\rm rad}$ can be written as
\begin{equation}
\rho_{\rm rad}=\frac{\pi^2}{30}g_{*}(T)\,T^4
\label{eqn10}
\end{equation}
If we consider the equation of state for the $\phi$ field to be $p_{\phi}=\omega_{\phi} \rho_{\phi}$ then Friedmann equation leads to $\rho_{\phi}\propto a^{-3(1+\omega_{\phi})}$. Therefore, one can connect $\omega_{\phi}$ and $n$ by the relation $n=3\omega_{\phi}-1$. Here, we always consider $n>0$, which implies that the $\phi$ energy always dominates over the radiation at early enough epochs. To express the energy density $\rho_{\phi}$ in terms of the radiation temperature, we consider that this new field $\phi$ does not have any interactions with SM particles and hence it only contribute to the energy density of the universe but not to the entropy density of the universe. This leads to the conservation of entropy in a comoving volume $S=sa^{3}=$constant, where the entropy density reads the standard one,
\begin{equation} \label{eq:52}
s(T)=\dfrac{2\pi^{2}}{45}g_{*s}(T)T^{3},
\end{equation}
with $g_{*s}$ being the effective relativistic degrees of freedom contributing to the entropy density. Taking the BBN constraints into account we argue that the equality between the energy density of $\phi$ and radiation must happen at a temperature $T_{\rm r}\gtrsim T_{\rm BBN}$. From equation \eqref{eq:52} and scaling the equation \eqref{feu1eq} one can write $\rho_{\phi}$ in terms of temperature as
\begin{equation}
\rho_{\phi}(T)=\rho_{\phi}(T_{\rm r})\left( \dfrac{g_{*s}(T)}{g_{*s}(T_{\rm r})} \right)^{(4+n)/3} \left( \dfrac{T}{T_{\rm r}} \right)^{4+n}.
\end{equation}
Then the full energy density at any temperature reads

\begin{equation} \label{eq:54}
\rho(T)=\rho_{\rm rad}(T)+\rho_{\phi}(T)=\rho_{\rm rad}(T) \left[ 1+\dfrac{g_{*}(T_{r})}{g_{*}(T)} \left( \dfrac{g_{*s}(T)}{g_{*s}(T_{\rm r})} \right)^{(4+n)/3}  \left( \dfrac{T}{T_{\rm r}}  \right) ^{n} \right].
\end{equation}
Considering $g_{*s}(T)=g_{*}(T)$ for most of the history of the universe the Hubble parameter can be calculated to be 
\begin{equation} \label{eq:55}
H(T)\simeq \dfrac{\pi g_{*}^{1/2}(T)T^{2}}{3 \sqrt{10} M_{\rm Pl}} \left[ 1+ \left( \dfrac{g_{*}(T)}{g_{*}(T_{\rm r})} \right)^{(1+n)/3} \left( \dfrac{T}{T_{\rm r}} \right) ^{n} \right]^{1/2}.
\end{equation}

\subsection{Dark matter}
\begin{figure}
\begin{center}
\includegraphics[scale=.35]{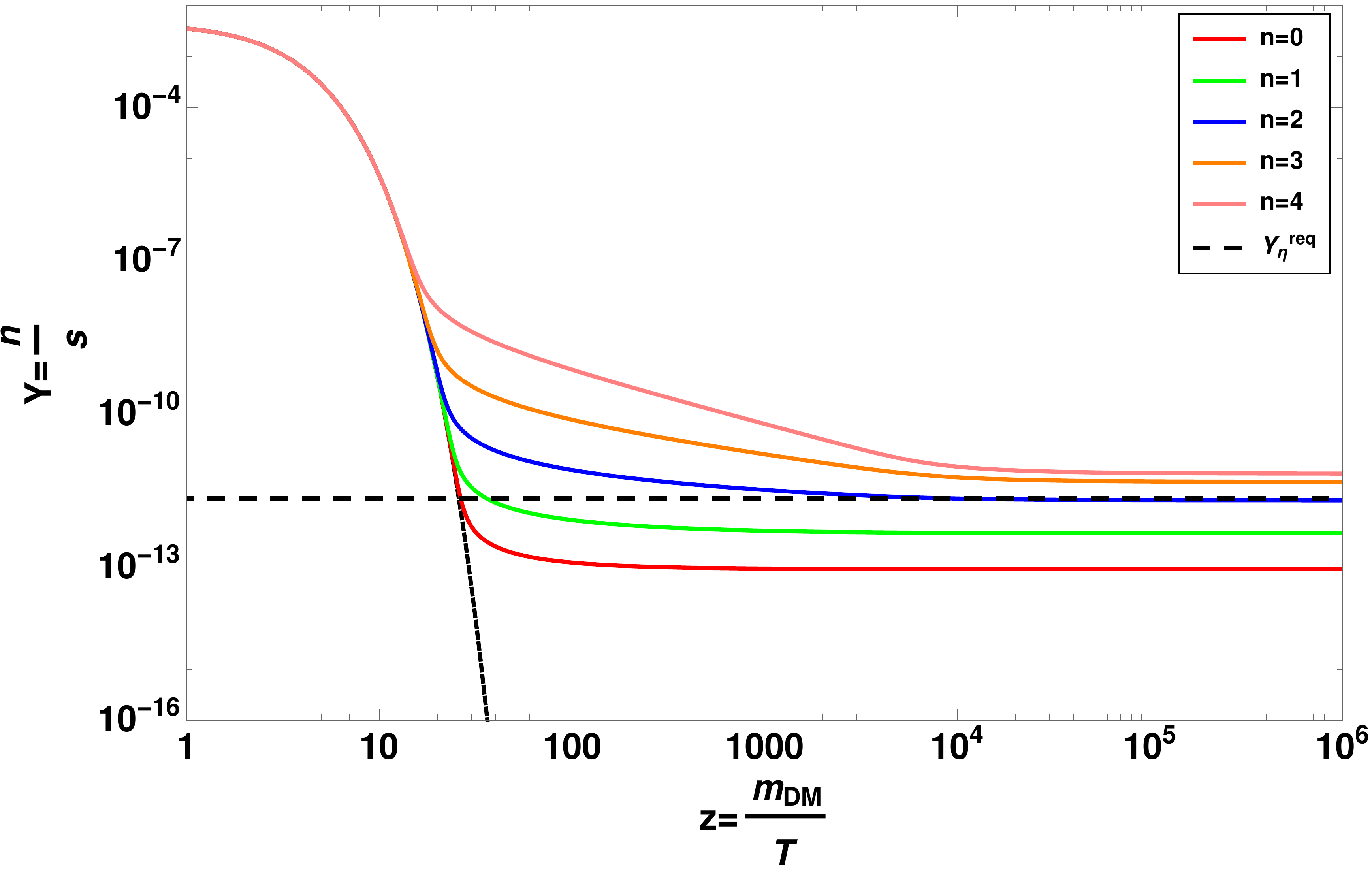} \label{Profumo}
\caption{Comoving number density of of DM ($\eta$) with $z$ for different cosmological histories. The parameters used for this results are $m_{DM}=200$ GeV, $m_{H_{0}}=200.076$ GeV ($\lambda_5=0.0005$), $m_{H^{\pm}}=205$ GeV, $T_{r}=20$ MeV and $\lambda_{L}=10^{-8}$. The black dashed line represents the required DM abundance with $200$ GeV mass to satisfy the correct Planck 2018 limit on DM abundance \cite{Aghanim:2018eyx}.}
\label{dmrelic:feu}
\end{center}
\end{figure}

In such a FEU scenario the Boltzmann equation to calculate the abundance of thermally produced DM is derived to be \cite{DEramo:2017gpl}

\begin{equation} \label{eq:56}
\dfrac{dY}{dz}=-A \dfrac{ \langle \sigma v_{\rm rel} \rangle}{z^{3}L \left[n,z,z_{r} \right]}\left[ Y^{2}-Y_{\rm eq}^{2} \right] ,
\end{equation} 
where, $A=\dfrac{s(z=1)}{H_{\rm rad}(z=1)}=\dfrac{2\sqrt{2}\pi}{3\sqrt{5}}g_{*}^{1/2}m_{\rm DM} M_{\rm Pl}$ and the function $L \left[n,z,z_{r} \right]$ has the form 
\begin{equation}
L \left[n,z,z_{r} \right]=\left({n+4}\right)\left[\dfrac{1}{z^4}+\left(\dfrac{g_{*}(z)}{g_{*}(z_{r})} \right)^{(1+n)/3}\dfrac{z_{r}^n}{z^{n+4}} \right] ^{3/2}\left[ \dfrac{4}{z^5}+(4+n)\left( \dfrac{g_{*}(z)}{g_{*}(z_{r})}\right)^{(1+n)/3}\dfrac{ z_{r}^n}{z^{n+5}} \right]^{-1}
\end{equation}
In the limit $T\gg T_{\rm r}$ the Boltzmann equation reduces to 

\begin{equation} \label{eq:58}
\dfrac{dY}{dz}=-A\dfrac{\langle \sigma v_{\rm rel} \rangle}{ z^{2-n/2}z_{r}^{n/2} } \left[ Y^{2}-Y_{\rm eq}^{2} \right].
\end{equation}
Borrowing this basic setup from \cite{DEramo:2017gpl} we first apply it to scalar doublet DM in our model. Note that, we do not ignore the sub-dominant radiation part in the calculation and use the most general Boltzmann equations to calculate DM relic. Similar to the earlier scenario, here also the annihilation cross sections of DM $\langle \sigma v_{\rm rel} \rangle$ are evaluated using \texttt{micrOMEGAs} package \cite{Belanger:2013oya}. We first show the evolution of DM density in figure \ref{dmrelic:feu} by choosing some benchmark values of DM parameters and for different integral values of $n$. The overall behaviour matches with the model independent analysis of earlier work \cite{DEramo:2017gpl}. Clearly, with increase in the values of $n$, the final DM relic abundance increases. While for $n=0$ or the usual radiation dominated universe, the DM remains under-abundant, $n=2$ FEU scenario generates the correct DM abundance for same benchmark DM parameters. We therefore, choose DM parameters in such a way that the usual thermal relic remains under-abundant. This justifies the choice of DM mass 200 GeV which falls in the range of inert Higgs doublet DM mass $\in (80-550)$ GeV where thermal abundance in usual radiation dominated universe remains suppressed. Thus, FEU scenario provides another way of generating correct relic in this intermediate mass range of inert Higgs doublet DM. Other possibilities by invoking non-thermal contribution and multi-component DM scenario can be found in \cite{Borah:2017dfn}, \cite{Borah:2019aeq} respectively.

\subsection{Leptogenesis}
After applying the basic recipe of FEU to the specific DM model we have, we now proceed to derive the Boltzmann equations for leptogenesis. The Boltzmann equation for leptogenesis in this scenario can be written as 
\begin{equation} \label{eq:59}
\dfrac{dn_{N_{1}}}{dz}=D_{1}^{'}(n_{N_{1}}-n_{N_{1}}^{\rm eq})
\end{equation}
\begin{equation} \label{eq:60}
\dfrac{dn_{B-L}}{dz}=-\epsilon_{1}D_{1}^{'}(n_{N_{1}}-n_{N_{1}^{eq}})-W_{\rm Total}^{'}n_{B-L}
\end{equation}
with the $z=M_1/T$ dependent quantities
\begin{equation} \label{eq:61}
D_{1}^{'}=K_{1}\dfrac{\kappa_{1}(z)}{\kappa_{2}(z)}\dfrac{1}{L\left[n,z,z_{r} \right]} ,
\end{equation}
\begin{equation} \label{eq:62}
W_{\rm Total}^{'}=W_{1}^{'}+\Delta W^{'} ,
\end{equation}
\begin{equation}
W_{1}^{'}=\dfrac{1}{4}z^{2}K_{1} \kappa_{1}(z)\dfrac{1}{L\left[n,z,z_{r} \right]} ,
\end{equation}
\begin{equation}
n_{N_{1}}=\dfrac{z^{2}}{2}\kappa_{2}(z)
\end{equation}
with the $\Delta W^{'}$ term taking care about the washouts coming from the $\Delta L=2$ scattering processes, $l\eta \longleftrightarrow \bar{l}\eta^{*}$ and $ll \longleftrightarrow \eta^{*} \eta^{*}$. This term can be calculated to be 

\begin{equation} \label{eq:65}
\Delta W^{'}=\dfrac{36\sqrt{5}M_{\rm Pl}}{\pi^{1/2}g_{l}\sqrt{g_{*}}v^{4}}\dfrac{1}{z^{3}L\left[n,z,z_{r} \right]}\dfrac{1}{\lambda_{5}^{2}}M_{1}\bar{m_{\zeta}}^{2}.
\end{equation}
Here $K_{1}$ is the usual decay parameter defined by $K_{1}=\dfrac{\Gamma_{1}}{H_{\rm rad}(z=1)}$. Other parameters are defined in a way similar to the usual leptogenesis in scotogenic model discussed in section \label{sec4}.
\begin{figure} \label{fig:19}
\begin{center}
\includegraphics[scale=.18]{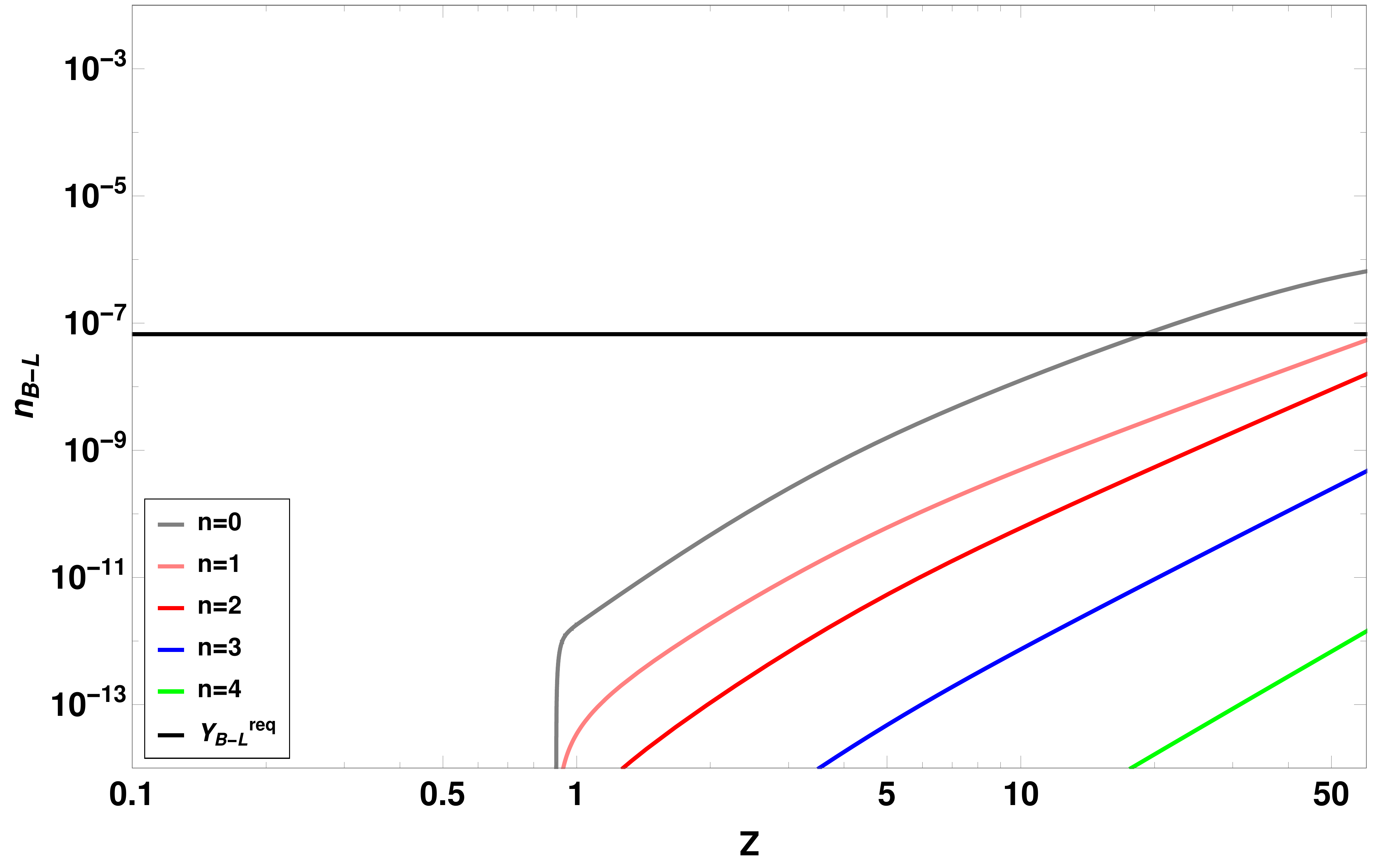}
\includegraphics[scale=.18]{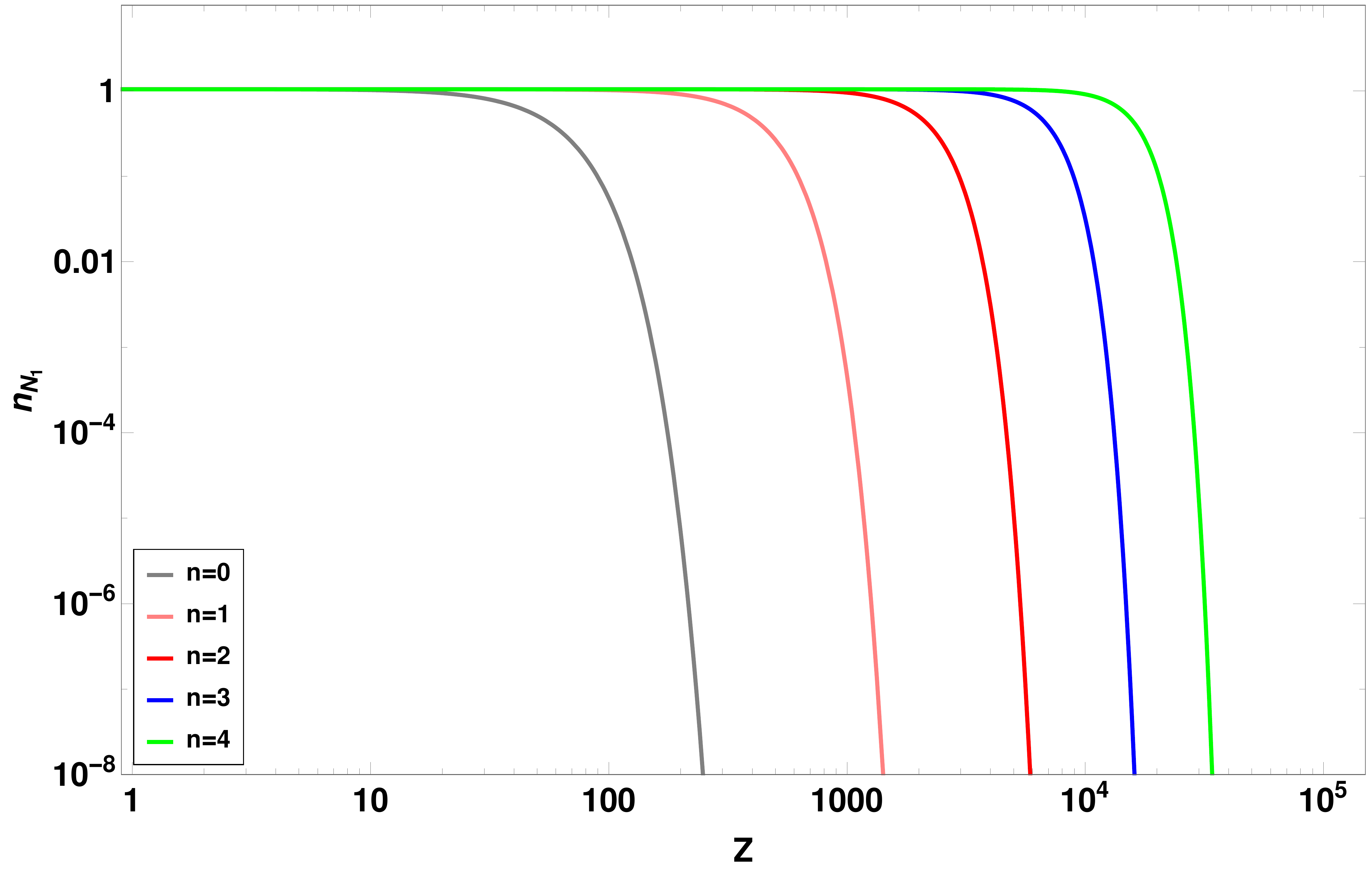}
\caption{ Evolution of comoving number density of $B-L$ with $z$ for different values of $n$ for FEU scenario (left panel), and evolution of the comoving number density of $N_{1}$ with $z$ for different values of $n$ for FEU scenario (right panel). Here we choose the parameters $m_{1}=10^{-13}$ eV, $M_{1}=10^{4}$ GeV, $M_{i+1}/M_{i}=10^{0.5}$ and $\lambda_{5}=10^{-4}$ and $T_{\rm r}=20$ MeV.}
\label{evol:feu1}
\end{center}
\end{figure}

\begin{figure}
\begin{center}
\includegraphics[scale=.30]{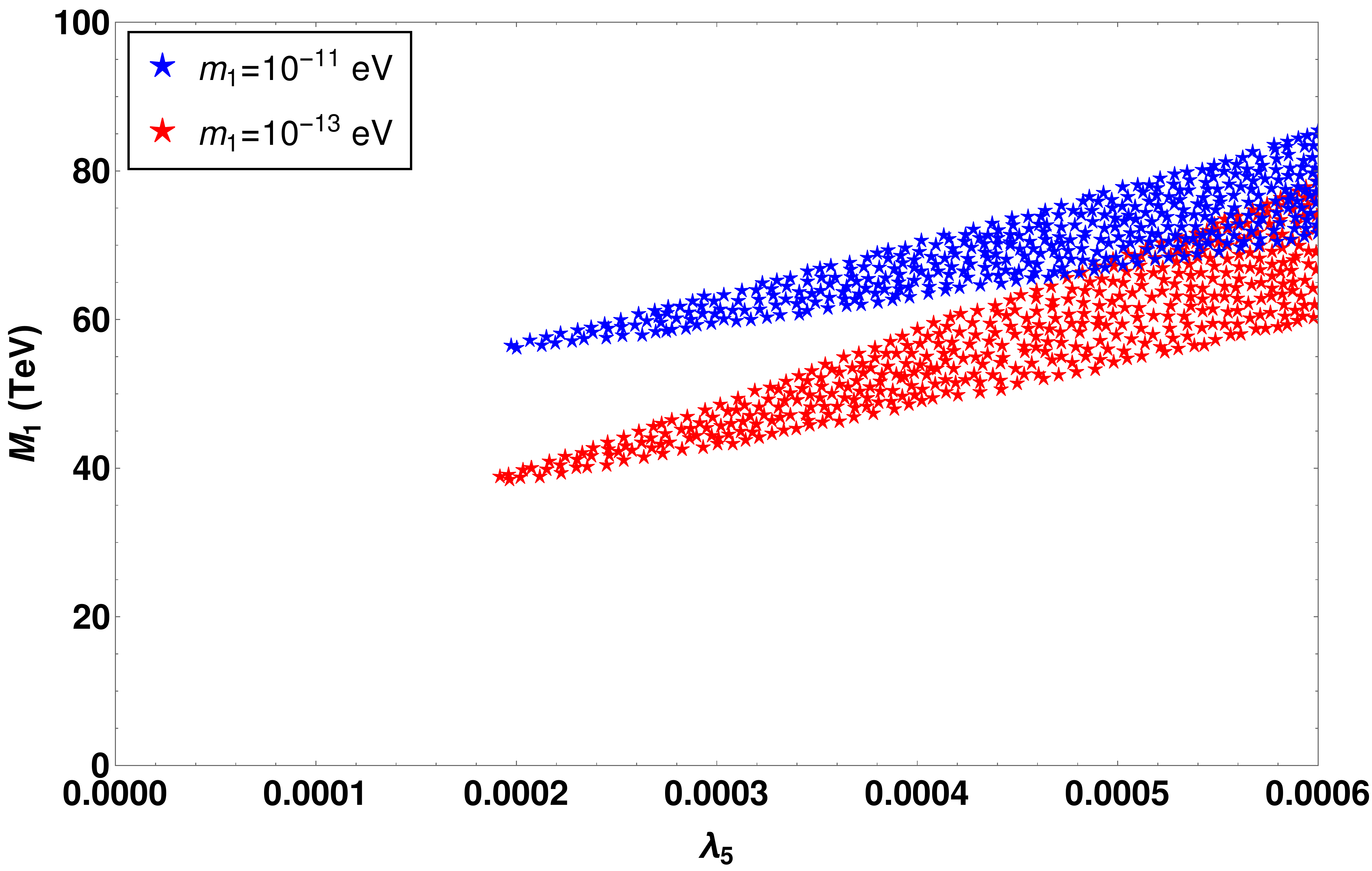}
\caption{Parameter space in the $M_{1}-\lambda_{5}$ plane giving rise to observed baryon asymmetry. The parameters used for this results are $\mu_2=200$ GeV, $n=2$,  $M_{i+1}/M_{i}=10^{0.5}$ and $T_{\rm r}=20$ MeV.}
\label{scan:feu1}
\end{center}
\end{figure}

Using the equations derived above, we first show the evolution of lepton asymmetry and $N_1$ abundance for different values of $n$ in figure \ref{evol:feu1} by choosing some benchmark values of model parameters. Clearly, there is a delay in generation of asymmetry as well as depletion in $N_1$ abundance with increase in the values of $n$. While $n=0$ or the standard cosmological scenario overproduces lepton asymmetry, the $n=1$ scenario produces the correct asymmetry as seen from left panel plot of figure \ref{evol:feu1}. In order to see the overall parameter space allowed from the requirement of generating correct baryon asymmetry, we fix $n=2$ and perform a numerical scan over $M_1-\lambda_5$ by keeping other model parameters fixed. The resulting parameter space is shown in figure \ref{scan:feu1}. While we are not showing scans for other possible values of $n$ here, the overall behaviour is similar and the scale of leptogenesis gets pushed up by a factor of order one compared to the standard cosmological scenario discussed earlier. A more rigorous numerical scan can be performed to find the complete parameter space that can generate correct baryon asymmetry of the universe.

\begin{figure}[h]
\includegraphics[scale=.3]{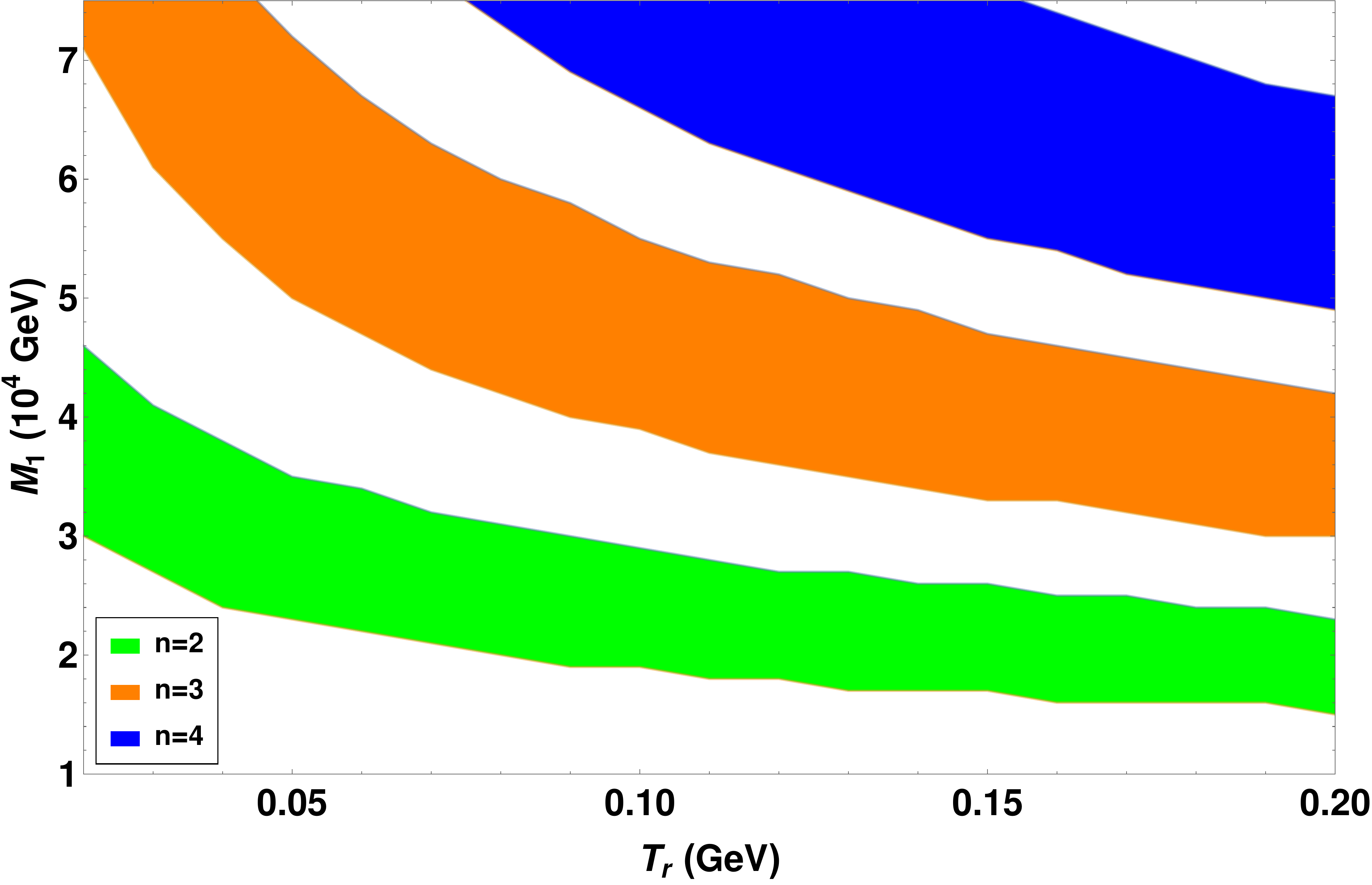}
\caption{Parameter space in $M_{1}-T_{\rm r}$ plane with different values of $n$ for successful leptogenesis in FEU scenario. The parameters used for this result are $\lambda_{5}=0.0003$, $\mu_{2}=200 $ GeV and $m_{1}=10^{-13}$ eV.}
\label{scan:feu1}
\end{figure}

We then constrain the cosmological parameters in FEU scenario from the requirement of producing the correct baryon asymmetry. We show it in $M_{1}-T_{\rm r}$ plane for different values of $n$ in figure \ref{scan:feu1}. We can see from this plot that for larger values $n$, we require large $M_{1}$ to satisfy the correct baryon asymmetry. Again for non-zero $n$,  large $T_{\rm r}$ pushes the cosmological history at and below the scale of leptogenesis more towards usual radiation like, bringing down the scale of leptogenesis (as in the standard case discussed earlier). 
 \begin{figure}[h]
\includegraphics[scale=.3]{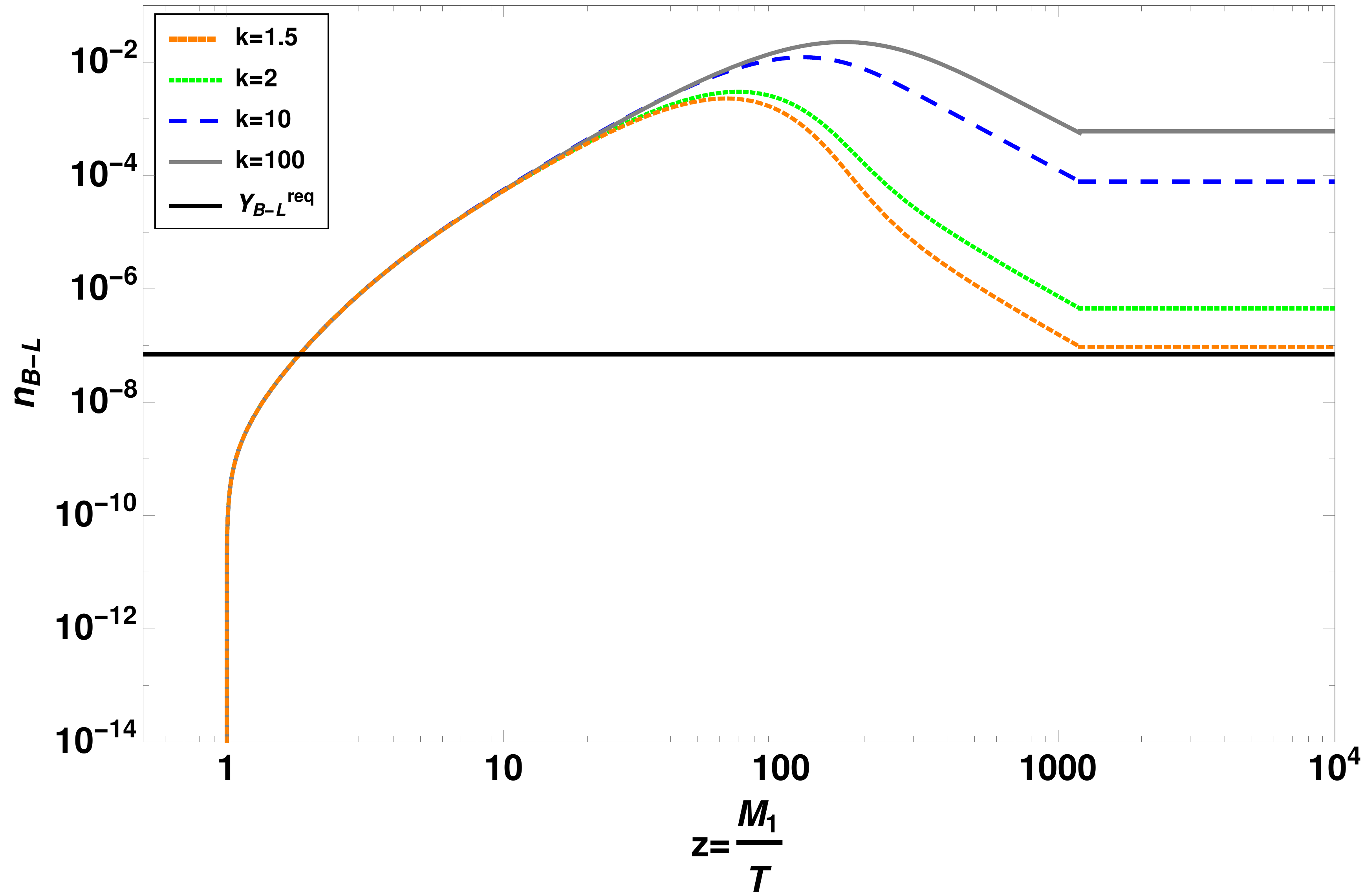}
\caption{Evolution of $n_{B-L}$  with $z=\dfrac{M_{1}}{T}$ in EMD universe (Case 2) including lepton flavour effects. The parameters used for these plots are $m_{1}=10^{-13}$ eV, $M_{1}=3\times10^{5}$ GeV, $M_{i}/M_{i+1}=10^{0.5}$, $\mu_2=3500$ GeV, $\lambda_{5}=10^{-4}$ and $T_{\rm end}=250$ GeV.}
\label{case2flav}
\end{figure}

\begin{figure}[h]
\includegraphics[scale=.3]{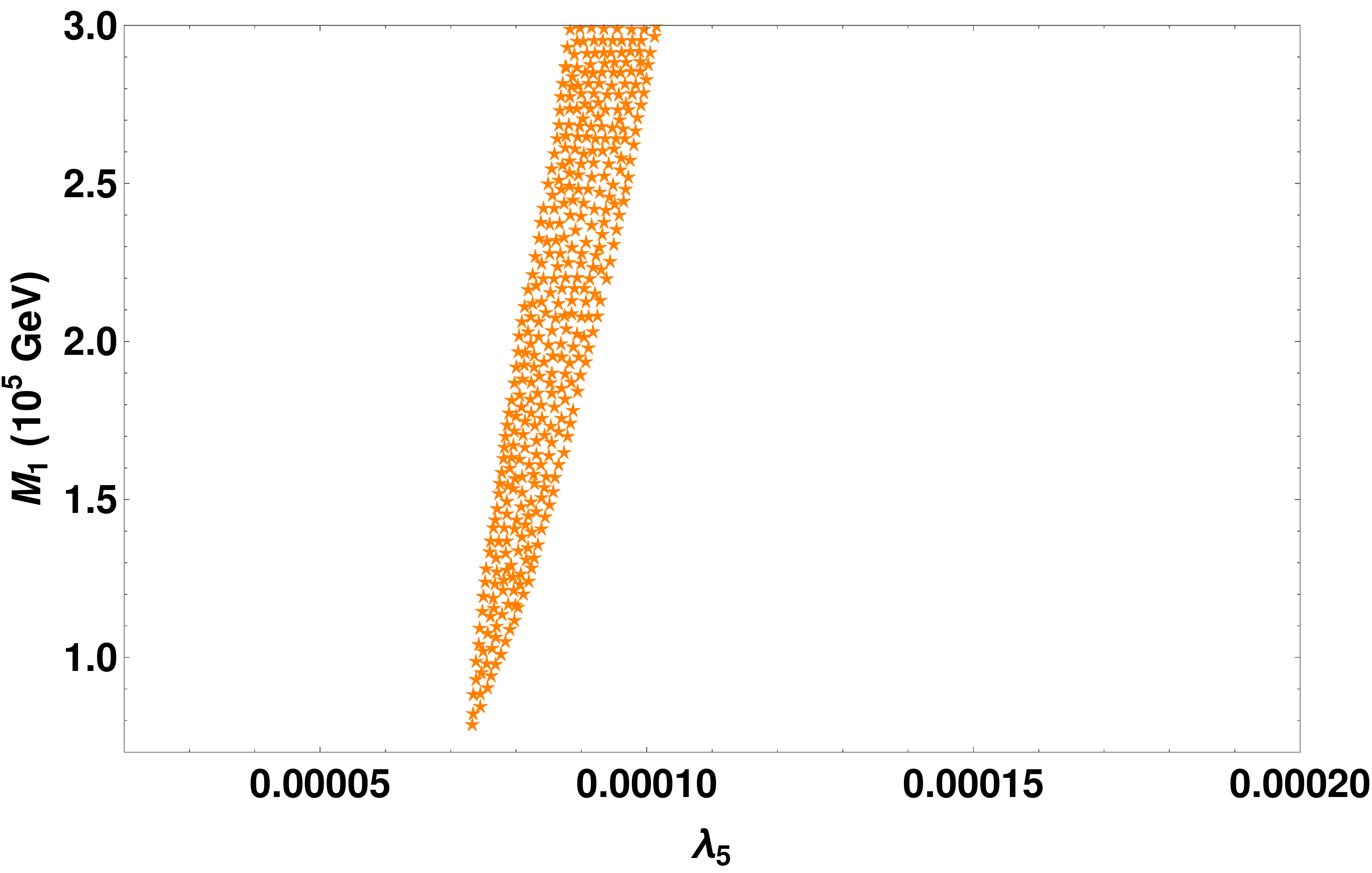}

\caption{Parameter space in $M_{1}-\lambda_{5}$ plane giving rise to the observed baryon asymmetry in EMD universe (Case 2) including lepton flavour effects. The benchmark parameters used for this result are $\mu_2=3500$ GeV, $m_{1}=10^{-13}$ eV and $M_{i+1}/M_{i}=10^{0.5}$. The cosmological parameters used for these results are $k=1.5$ and $T_{\rm end}=250$ GeV.}
\label{case2scanflav}
\end{figure}
%\subsubsection{Case 3: $T_{end}<<T_{Sphaleron}<<T_{eq}$}

\begin{figure}[h]
\includegraphics[scale=.22]{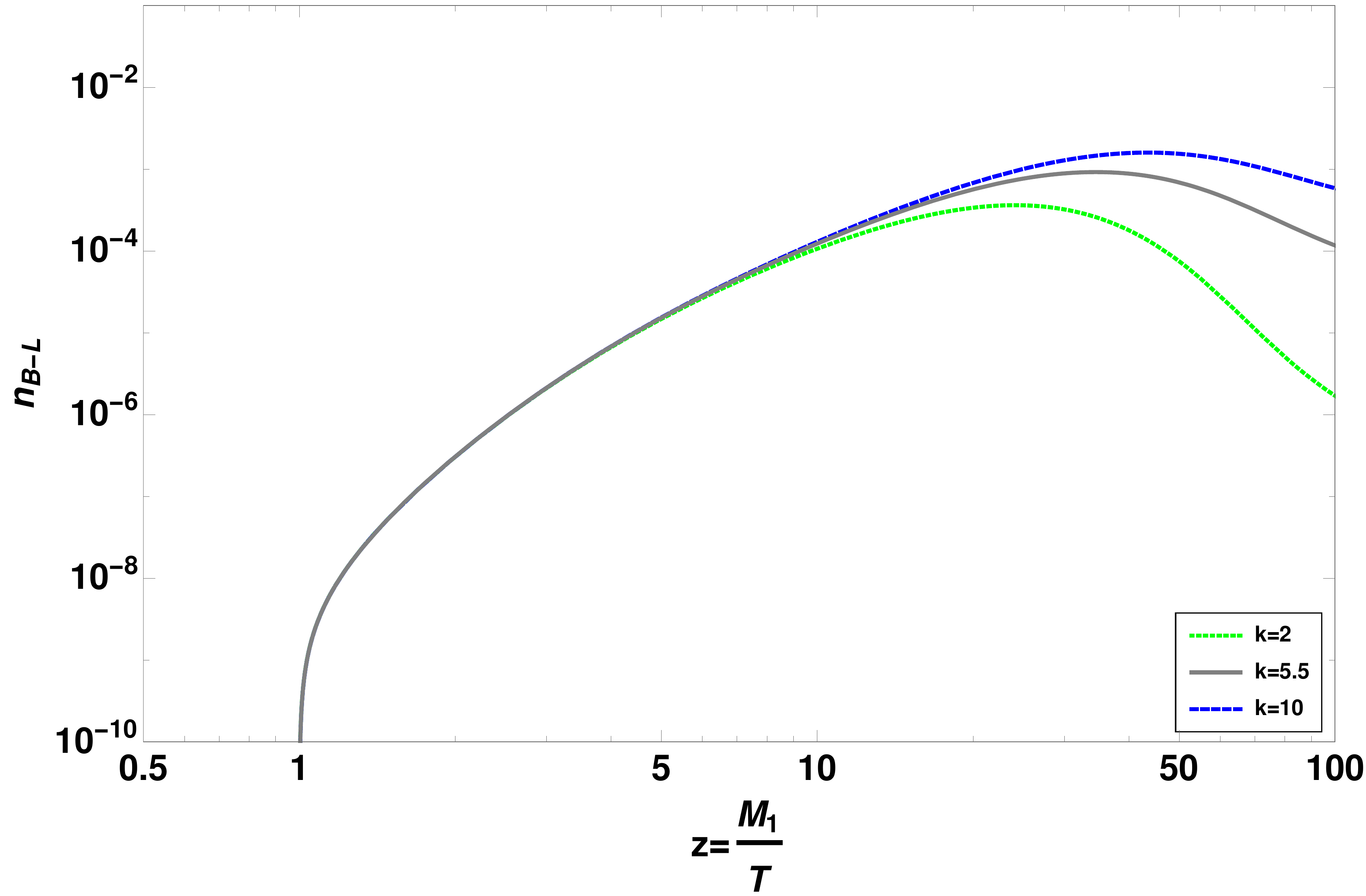}
\includegraphics[scale=.22]{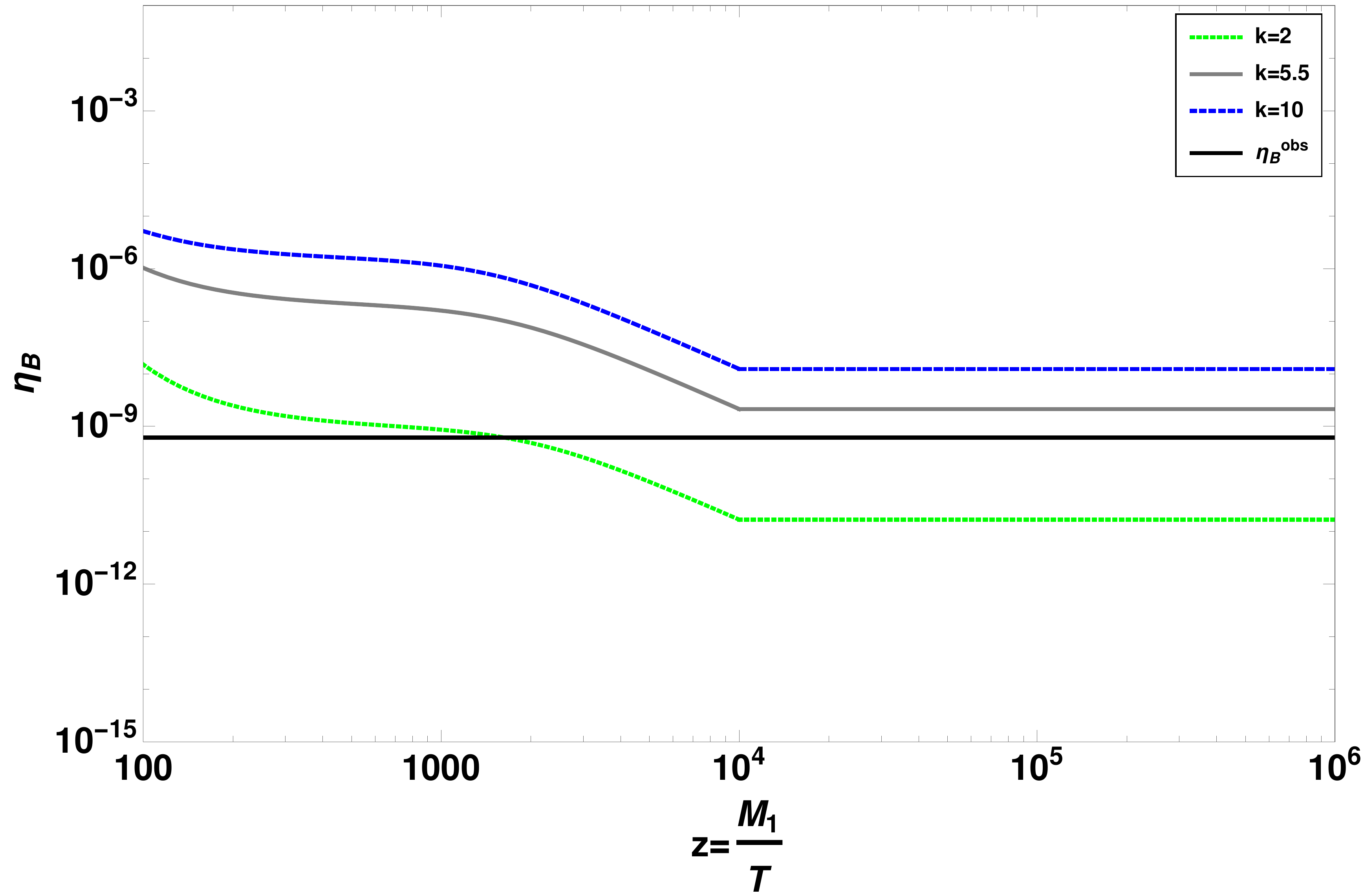}
\caption{Evolution of $n_{B-L}$ and $\eta_{B}$ with $z=\dfrac{M_{1}}{T}$ in EMD universe (Case 3) including lepton flavour effects. The parameters used for these plots are $m_{1}=10^{-13}$ eV, $M_{1}=3\times10^{3}$ GeV, $M_{i}/M_{i+1}=10^{0.5}$, $\mu_2=1000$ GeV, $\lambda_{5}=10^{-4}$ and $T_{\rm end}=1$ GeV.}
\label{case3flav}
\end{figure}

\begin{figure}[h]
\includegraphics[scale=.3]{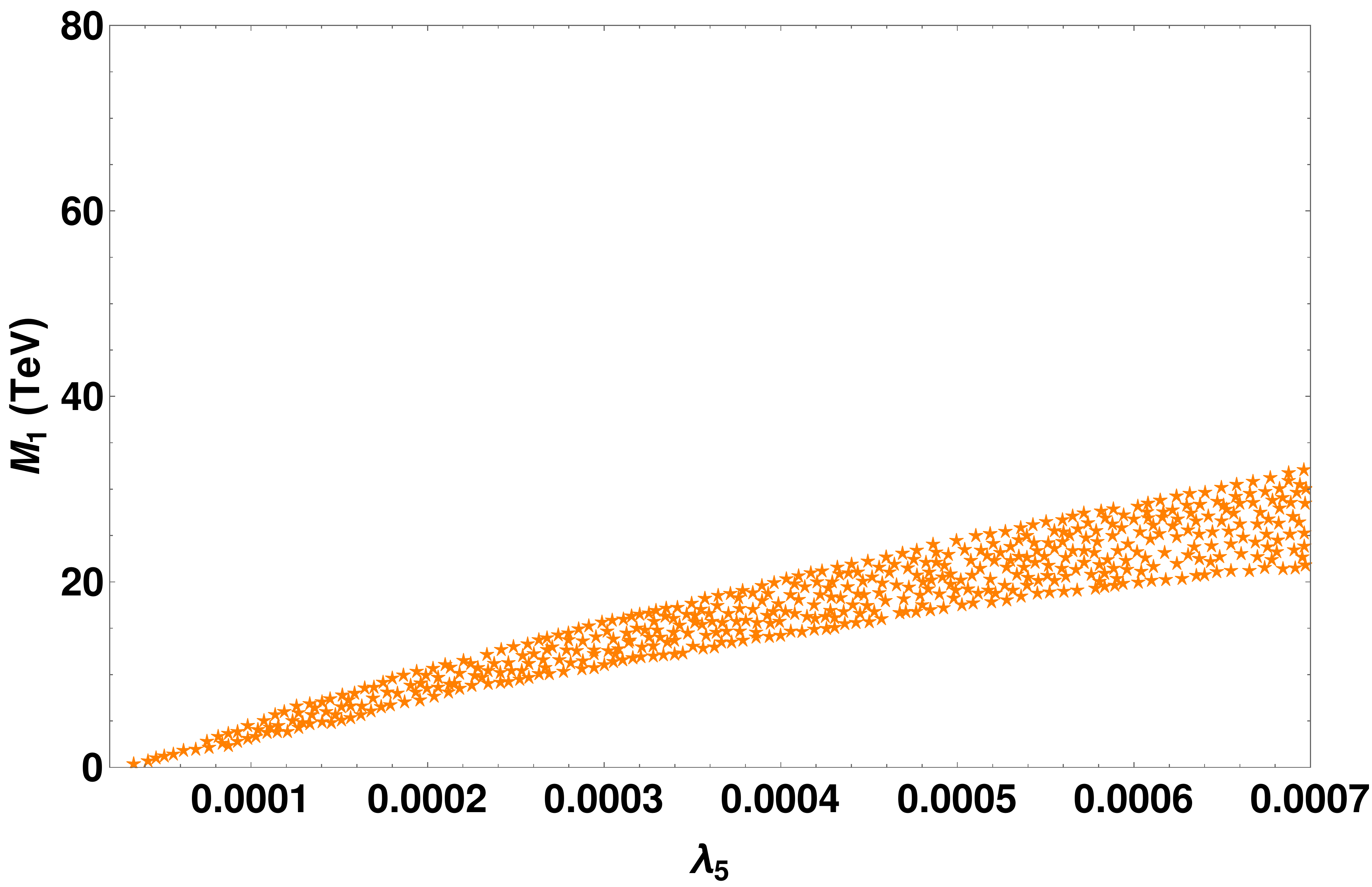}
\caption{Parameter space in $M_{1}-\lambda_{5}$ plane giving rise to the observed baryon asymmetry in EMD universe (Case 3) including lepton flavour effects. The benchmark parameters used for this result are $\mu_2=1000$ GeV, $m_{1}=10^{-13}$ eV and $M_{i+1}/M_{i}=10^{0.5}$. The cosmological parameters used for these results are $k=5.5$ and $T_{\rm end}=1$ GeV.}
\label{case3scanflav}
\end{figure}
\begin{figure}[h]
\includegraphics[scale=.25]{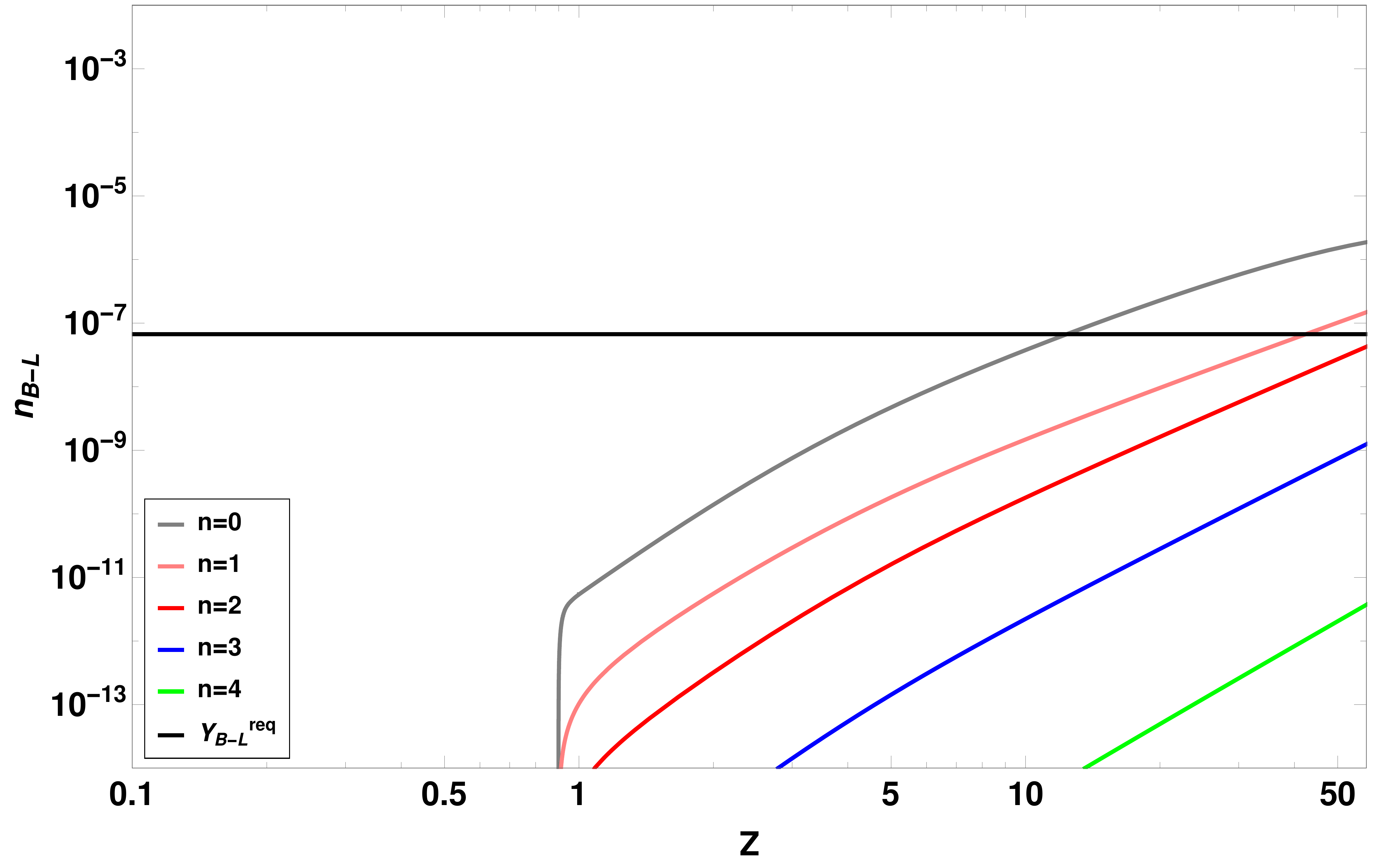}
\caption{Evolution of comoving number density of $B-L$ with $z$ for different cosmological models with $z$ for different cosmological models. Here we choose the parametrs $m_{l}=10^{-13}$ eV, $M_{1}=10^{4}$, $M_{i+1}/M_{i}=10^{0.5}$ and $\lambda_{5}=10^{-4}$ and $T_{r}=20$ MeV.}
\label{feuflav}
\end{figure}
\begin{figure}[h]
\includegraphics[scale=.25]{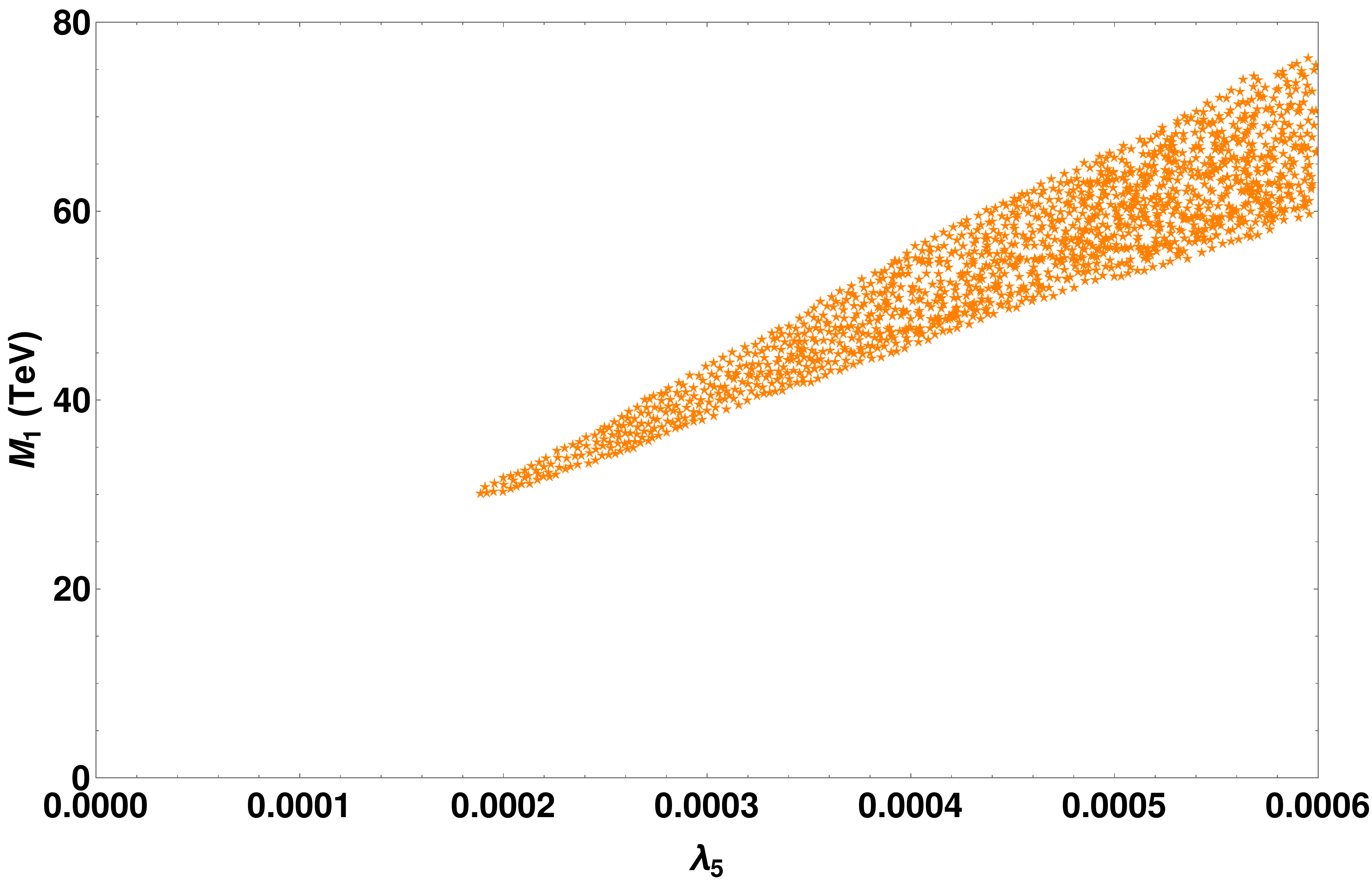}
\caption{Parameter space in $M_{1}-\lambda_{5}$ plane giving rise to the observed baryon asymmetry in FEU scenario including lepton flavour effects. The benchmark parameters used for this results are $\mu_{2}=200$ GeV, $M_{i+1}/M_{i}=10^{0.5}$ and $m_{1}=10^{-13}$ eV. The cosmological parameters used for these results are $n=2$ and $T_{\rm r}=20$ MeV.}
\label{feuscanflav}
\end{figure}

\section{Leptogenesis with Flavour effects}
\label{sec:flav}
In the discussions above, we have not considered the effects of lepton flavours. In this section, we briefly summarise the role of lepton flavour effects on our results. For earlier works on flavoured leptogenesis, please refer to \cite{Abada:2006fw, Abada:2006ea, Nardi:2006fx, Blanchet:2006be} while a recent review may be found in \cite{Dev:2017trv}. Here we adopt the notations adopted by \cite{Blanchet:2006be} and calculate the baryon asymmetry for all the non-standard cosmological scenarios mentioned earlier, including lepton flavour effects.
%\subsection{Flavoured Leptogenesis in early matter dominated universe}

The Boltzmann equations for flavoured leptogenesis in EMD universe can be written as,
\begin{equation}
\dfrac{dn_{N_{1}}}{dz}+\dfrac{n_{N_{1}}}{s}\dfrac{ds}{dz}+\dfrac{3n_{N_{1}}}{z}=-D(n_{N_{1}}-n_{N_{1}}^{\rm eq})
\end{equation}
\begin{multline*}
\dfrac{dn_{\Delta _{\alpha}}}{dz}+\dfrac{n_{\Delta _{\alpha}}}{s}\dfrac{ds}{dz}+\dfrac{3n_{\Delta_{\alpha}}}{z}=-\epsilon_{1\alpha}D(n_{N_{1}}-n_{N_{1}}^{\rm eq})-P_{1\alpha}W_{1}n_{\Delta_{\alpha}}-P_{1\alpha} n_{\Delta \alpha} \sum_{\beta=e,\mu,\tau}P_{1\beta} (\Delta W)_{\alpha \beta}.
\end{multline*}
Here $\alpha=e,\mu,\tau$ corresponds to lepton flavours resulting in three coupled equations. $n_{\Delta \alpha}$ is the comoving number density of $B/3-L_{\alpha}$ (for each flavour of leptons) and $P_{1\alpha}$ are the projectors defined by 
  \begin{equation}
 P_{1\alpha}=\dfrac{\Gamma_{1\alpha}}{\Gamma_{1}},
 \end{equation}
 where $\Gamma_1$ is the total decay width of $N_1$ while $\Gamma_{1\alpha}$ is the corresponding partial decay width to a particular lepton flavour denoted by $\alpha$. The flavoured CP asymmetry parameter $\epsilon_{1\alpha}$ and the washout terms can be found in a way similar to the unflavoured leptogenesis discussed earlier. In EMD scenario, they are solved simultaneously with equations \eqref{eq:71}, \eqref{eq:72} like before. The corresponding evolution of asymmetry for EMD case 2: $T_{\rm sphaleron}\lesssim T_{\rm end} \ll T_{\rm eq}$ is shown in figure \ref{case2flav} which is similar to the unflavoured regime. The allowed parameter space in $M_{1}-\lambda_{5}$ plane is shown in figure \ref{case2scanflav}. The corresponding plots for EMD case 3: $T_{\rm end} \ll T_{\rm sphaleron} \ll T_{\rm eq}$ are shown in figure \ref{case3flav} and \ref{case3scanflav}. Comparison with the results in unflavoured case indicates that the inclusion of lepton flavour effects slightly lowers the scale of leptogenesis, as expected. It should be noted that while comparing flavoured leptogenesis with unflavoured ones (that is, comparing figure \ref{case3scanflav} with figure \ref{fig:3emda}), we have used the lightest active neutrino mass to be $m_1 = 10^{-13}$ eV which corresponds to the weak washout regime $K_1 = \Gamma_1/H(z=1) \ll 1$ \cite{Blanchet:2006be}. In such weak washout regimes the projection operator $P_{1\alpha} \rightarrow 1$ taking the flavoured leptogenesis towards the unflavoured regime. Therefore, the quantitative difference after inclusion of flavour effects in this weak wash-out regime is less compared to the difference brought out by non-standard cosmological epoch (EMD) over standard cosmology. Going to strong washout regime will show the significance of flavour effects in a clear way. However, since the scale of leptogenesis is pushed up in such a scenario even with non-standard cosmology, we do not discuss this any further.

%\subsubsection{Flavoured Leptogenesis in fast expanding universe}
Similarly, the Boltzmann equations for flavoured leptogenesis in fast expanding universe scenario can be written as
\begin{equation}
\dfrac{dn_{N_{1}}}{dz}=-D_{1}^{'}(n_{N_{1}}-n_{N_{1}}^{\rm eq})
\end{equation}
\begin{equation}
\dfrac{dn_{\Delta _{\alpha}}}{dz}=-\epsilon_{1\alpha}D_{1}^{'}(n_{N_{1}}-n_{N_{1}}^{\rm eq})-P_{1\alpha}W_{1}^{'}n_{\Delta_{\alpha}}-P_{1\alpha} n_{\Delta \alpha} \sum_{\beta=e,\mu,\tau}P_{1\beta} (\Delta W^{'})_{\alpha \beta}.
\end{equation}
The notations have their usual meaning as adopted throughout the discussions. The evolution of lepton asymmetry for flavoured leptogenesis in FEU scenario is shown in figure \ref{feuflav} while the allowed parameter space in $M_{1}-\lambda_{5}$ plane is shown in figure \ref{feuscanflav}. While the evolution of asymmetry looks very similar to the unflavoured case, the allowed parameter space reveals that the scale of leptogenesis can be lower by a numerical factor of order one compared to the unflavoured case.

\section{Conclusion}
\label{sec:conc}
We have studied the possibility of generating correct baryon asymmetry and dark matter relic in the universe by considering two different non-standard cosmological epochs prior to the BBN era. Considering the TeV scale minimal scotogenic model which generates light neutrino masses at one loop, we consider the lightest $Z_2$ odd particle (assumed to be one of the neutral components of a scalar doublet) to be the DM candidate while the out-of-equilibrium decay of heavy singlet neutrinos to be the origin of baryon asymmetry via leptogenesis. While DM relic calculation in such non-standard cosmological epochs were studied earlier by several authors, the detailed calculations for leptogenesis were missing. Apart from filling this gap, our motivation has been to check if such non-standard cosmological epochs can help us to lower the scale of leptogenesis compared to the standard radiation dominated scenario. Also, the DM relic calculation recipe in such non-standard cosmological scenarios is applied to a very popular and specific particle physics model in our work, in contrast with model independent approaches adopted in earlier works. Alternately, such a study can also constrain such non-standard cosmological scenarios from the requirement of generating correct baryon asymmetry of the universe along with DM relic while being consistent with other phenomenological requirements like light neutrino mass and direct search bounds.

In the first non-standard cosmological scenario where we assumed an early matter dominated epoch, we found one possible realisations where the scale of leptogenesis can be as low as 1 TeV, significantly lower than the scale $\mathcal{O}(10 \;\rm TeV)$ found in the minimal scotogenic model by previous studies while considering a radiation dominated universe. We also point out other possible realisations which are either trivial (due to similarity with usual standard cosmological scenario) or inconsistent with correct leptogenesis at low scale. In the second non-standard cosmological scenario, we consider a scalar field to dominate the energy density of the universe such that the energy density of the scalar field red-shifts faster than usual radiation, known as fast expanding universe. We found that in this case the leptogenesis scale gets pushed slightly higher compared to the standard case. We found that successful leptogenesis can be achieved with singlet neutrino mass as low as $M_{1}\simeq 40$ TeV also satisfying the dark matter relic with $m_{\rm DM}=200$ GeV for the case $\rho_{\phi}\propto \dfrac{1}{a^{6}}$ or $n=2$. Inclusion of lepton flavour effects can further lower the scale of leptogenesis by a numerical factor, as we point out towards the end of our analysis. All such scenarios we study here are also consistent with light neutrino masses and mixing. While our numerical analysis were confined to scan over a limited number of parameters, a more detailed and rigorous numerical scan of such non-standard cosmological scenarios should lead to more interesting possibilities. We leave such a detailed calculation to future works. UV complete realisations of the non-standard cosmological phases and particle physics implications are also left for an upcoming work.

The TeV scale leptogenesis and DM with non-standard cosmological history can, in principle, give rise to interesting observational signatures. Direct probe of leptogenesis is a difficult task, specially in the right handed neutrino decay framework. However, there exists ways to falsify high scale leptogenesis by observations of lepton number violating low energy effects \cite{Chun:2017spz}. Since the scale of leptogenesis gets lower in one of the non-standard cosmological epochs of early matter domination, it will be interesting study the possibility of falsifying it at ongoing experiments. Probing this EMD epochs from signatures related to DM gets difficult as the requirement of thermal overproduction (to compensate for late entropy dilution) pushes the DM mass into heavier regimes. On the other hand, non-standard cosmological epoch can leave imprint on gravitational waves to be probed by upcoming experiments \cite{Boyle:2007zx, Cui:2018rwi}, can also affect the rate of expansion of the universe providing a solution to the discrepancy between high and low redshift measurements of the Hubble parameter \cite{Knox:2019rjx}. In the fast expanding universe scenario, although the scale of leptogenesis is pushed towards higher side, possibility of DM mass in the intermediate mass regime of scalar doublet DM ($\in (80-550)$ GeV) offers interesting probe. As noted in \cite{Borah:2017dfn}, DM in this mass range has a large annihilation rate into $W$ boson pairs that can give rise to excess of gamma rays which can be probed by Fermi-LAT \cite{TheFermi-LAT:2017vmf}, \cite{HESS:2015cda}. Additionally, such new low mass regime satisfying relic can also give rise to collider signatures which were not considered previously. We leave such observational aspects to future studies.

\acknowledgments
DB acknowledges the support from Early Career Research Award from the department of science and technology-science and engineering research board (DST-SERB), Government of India (reference number: ECR/2017/001873) and Associateship Programme of Inter University Centre for Astronomy and Astrophysics (IUCAA), Pune.

%\newpage
%\bibliographystyle{JHEP}
%\bibliography{ref}
\providecommand{\href}[2]{#2}\begingroup\raggedright\endgroup

\end{document}